# WAVE-PARTICLE DUALITY REVITALIZED: CONSEQUENCES, APPLICATIONS AND RELATIVISTIC QUANTUM MECHANICS

**Himanshu Chauhan, Swati Rawal and R K Sinha**
*TIFAC-Centre of Relevance and Excellence in Fiber Optics and Optical Communication, Department of Applied Physics, Delhi Technological University (Formerly Delhi College Of Engineering, University of Delhi), Bawana Road, Delhi-110042, India*
E. Mail : *chauhan.himanshu_om@yahoo.com,*
*swati.rawal@yahoo.com,*
*dr_rk_sinha@yahoo.com*

**PACS numbers**: - 03.65-Ta, 03.65Ge, 03.65Pm, 03.75-b

## ABSTRACT

The proposed paper presents the unobserved inadequacies in de Broglie's given concepts of wave-particle duality and matter waves in the year 1923. The commonly admitted quantum energy or frequency expression $h\nu=\gamma mc^2$ is shown to be inappropriate for matter waves and is acceptable only for photons, where the symbols have their usual meanings. The superluminal phase velocity expression $c^2/\upsilon$, for matter waves, is investigated in detail and is also reported to be inadequate in the proposed paper. The rectifications in the inadequate concepts of de Broglie's theory and refinements in the analogy implementation between light waves and matter waves are presented, which provides the modified frequency and phase velocity expression for matter waves. Mathematical proofs for the proposed modified frequency and phase velocity expression are also presented. In accordance with the proposed concepts, a wave-particle duality picture is presented which elucidates the questions coupled with the wave-particle duality concepts, existing in the literature. Consequently, particle type nature is shown to be a characteristic of waves only, independent from the presence of matter. The modifications introduced in the frequency expression for matter waves leads to variation in the wave function expression for a freely moving particle and its energy operators, with appropriate justifications provided in the paper. A new relation between the Kinetic energy and Momentum of the moving body is also proposed and is subsequently applied to introduce novel General and Relativistic Quantum Mechanical Wave Equations. Applications of these equations in bound state quantum mechanical systems, presented in the paper, provide the information regarding particle's general and relativistic behavior in such systems. Moreover, the proposed wave equations can also be transformed into Schrödinger's and Dirac's equations. The interrelation of Schrödinger's, Dirac's and proposed equations with the universal wave equation is also presented.

## 1. INTRODUCTION

The history of science depicts that the reasonable ignorance of the well-established concepts has always been the starting-point in the development of a completely new branch of knowledge. M. Planck [1-4], A. Einstein [5-9], J.C. Maxwell [10] etc. are the well-known examples for such advancements. For the explanation of various phenomena of physics and to remove various inadequacies, they went across the boundaries of known science and proposed their revolutionary concepts against the traditional theories. The theory proposed by them, proved to be successful enough to investigate new domains of physics. J.C. Maxwell, around 1865 [10], uncovered the hidden inadequacies of Ampere's law and introduced the concept of displacement current for its rectification. Similarly, in the proposed paper, the inadequacies persisting in the wave-particle duality theory, proposed by Louis de Broglie [11-17], are uncovered and their rectifications are proposed. *The theory proposed is capable of providing a new insight in Quantum Mechanics.*

The proposal of light as an electro-magnetic wave [18], given by James Clerk Maxwell (1865) [10], explained various wavy phenomenon of light like interference and diffraction. However, in the beginning of the $20^{th}$ century, particle type nature of light also came into picture, with many experimental supports, e.g. Black body radiation phenomenon [1-4], Photo-electric effect [6,19-20], Compton effect [21,22], Raman effect [23-25] etc.. Thus current scientific theory holds that light has wave as well as particle nature [26]. In 1923, de Broglie hypothesized that the wave-particle duality should be common to both radiation and matter [11, 12,27], i.e. a moving body may also be associated with some waves, named matter waves [28,29].The phase velocity, group velocity, wavelength and frequency expressions for the matter waves proposed by de Broglie [11-17,28,29] are:-

$$\upsilon_P = \frac{c^2}{\upsilon}; \upsilon_g = \upsilon; \lambda = \frac{h}{\gamma m \upsilon} = \frac{h}{P}; E = h\nu = \gamma mc^2$$

where, $\upsilon_P$ is the phase velocity, $\upsilon$ is the velocity of the body, $\upsilon_g$ is the group velocity, $\lambda$ is the wavelength, $\nu$ is the frequency of matter waves .Simultaneously, $P$ is the momentum, $E$ is the *total energy*, and $m$ is the inertial mass of the body, $h$ is the Planck's constant and '$c$' is the speed of light in vacuum. The term $\gamma$ is defined as:-

$$\gamma = \frac{1}{\sqrt{1-\frac{\upsilon^2}{c^2}}}$$

The experiments carried out to observe the wave characteristics of matter [30-36], verified the matter wave's wavelength expression, proposed by de Broglie.

The proposed paper deals with the misconceptions of wave-particle duality theory. *The frequency and phase velocity expression for matter waves, proposed by de Broglie, is shown to be inadequate and the modified expressions for them are presented.* According to his proposed concept [11-17], the frequency of matter waves is related to the *total energy* ($E$) of body, i.e. de Broglie's proposed frequency expression:-

$$h\nu = \gamma mc^2 = \frac{mc^2}{\sqrt{1-(\upsilon/c)^2}}$$

This commonly admitted quantum energy or frequency expression, $h\nu=\gamma mc^2$, is revealed to be inappropriate for matter waves and is applicable only for photons. The superluminal phase velocity expression $c^2/\upsilon$, for matter waves, is investigated in detail and is also reported to be inadequate in the proposed paper. *Therefore, the modified expressions for both frequency and phase velocity are proposed.*

Antonis N. Agathangelidis [37] in his research paper attempted to eliminate the superluminal phase velocity concept of matter waves, by proposing that the matter waves should follow the moving body from behind, at a velocity which is half the non-relativistic velocity of the particle. However, the research paper by Antonis N. Agathangelidis [37] has several misconceptions. **First**, the assumption took by de-Broglie to have $(1/\lambda) = 0$ for $\upsilon=0$, where $\upsilon$ is the velocity of body, was disapproved by Agathangelidis, which according to him [37] is responsible to offer a wrong expression for the phase velocity of matter waves. However, this postulation is neither incorrect and nor responsible for the origin of phase velocity expression for matter waves. De Broglie had an extremely strong and concrete physical justification for the infinite wavelength of matter waves, corresponding to the body at rest. During his Nobel lecture (1929) [16], de Broglie elucidated that in the rest frame of the body, associated matter waves are also stationary; their phase is identical everywhere in this frame. Therefore, it becomes clear that the wavelength of matter waves should become infinite, when the body is immobile. Additionally, the experimentally verified wavelength expression [30-36], $\lambda=h/P$, articulates the infinite wavelength of matter waves for the zero velocity of body. Therefore, the infinite wavelength of matter waves, for zero velocity of body, becomes essentially evident. **Second**, the concept of group velocity of matter waves is kept unconcerned by Agathangelidis [37]. He treated matter waves as pure monochromatic phase waves, maintaining a velocity, which is half the non-relativistic velocity of body. However, Heisenberg's uncertainty principle, $\Delta E \Delta t \geq h/4\pi$ [27-28], elucidates the non-existence of a pure monochromatic wave; therefore, inclusion of the group wave concept is necessary. **Third**, Agathangelidis [37] regarded matter waves as separated from the body, tracking it from behind. However, this is not possible, because matter waves are responsible for the wavy nature of a body and therefore should always be associated with it. As a result, matter waves should always propagate in the company of the body. **Fourth**, the derivation of the relation $\lambda=h/P$, presented by Agathangelidis [37], is regarded as the derivation proposed by de Broglie. However, de Broglie offered the derivation and concepts in a completely different manner and are discussed in the proposed paper, in detail. **Fifth**, *Agathangelidis [37] has substituted the term hv, equivalent to the kinetic energy of body/system, without stating any reason. On the basis of this non-justified substitution, further derivation of the phase velocity expression for matter waves is carried out. However, not even a single specific argument for the substitution is stated.* He also criticized de-Broglie, by calling him as an etherist and opposed the experimentally verified wavelength relation, $\lambda=h/P$. **Sixth**, contrary to his rejection of the relation, $\lambda=h/P$, Agathangelidis simultaneously adopted it for the derivation of his modified phase velocity expression. **Seventh**, he considered matter waves as *slowly* moving waves which originates in the vibrations of ether medium, *after passage of the moving body* [37]. All these concepts given by him are inappropriate. The concept of ether has long been disproved mainly by Albert A. Michelson [38-40], by his famous 'Michelson-Morley experiment'.

In the present paper, detailed study of the concepts proposed by de Broglie is carried out and inadequacies are framed out. Rectifications for these inadequate concepts are proposed in the present paper. The frequency expression proposed by de Broglie for the matter waves is the primarily inadequate. The frequency and phase velocity expression, proposed by de Broglie, resulted from the inappropriate concepts presumed by him. Numerous additional self negations of the concepts are framed out from his research papers [11-13,15,17], thesis [14] and Nobel lecture [16]. The frequency expression proposed by de Broglie [11-17] for the matter waves is primarily inadequate. Many significant imperfections in the matter wave's frequency expression are brought up in the proposed paper, e.g. when the body is at rest, *de Broglie's proposed frequency expression articulates a finite non-zero value, despite of the fact that experimentally verified wavelength expression* $\lambda=h/\gamma m\upsilon$ *dictates the wavelength of matter waves to be infinite, which is in contrast to the finite frequency obtained from* $h\nu=\gamma mc^2$. Corresponding to the infinite wavelength of matter waves, the frequency should become zero, which opposes the finite frequency obtained from de Broglie's proposed frequency expression for matter waves ($h\nu=\gamma mc^2$). In view of the experimental verification for the wavelength expression [30-36], the inappropriateness of the frequency expression becomes self-evident.

The frequency expression is a result of the consideration of photons and the body to be analogous [11-17], thus for matter waves de Broglie proposed:-

$$h\nu = \gamma mc^2$$

One point to be clarified at this moment is that, it is not that the analogy between light waves and matter waves hypothesized by de Broglie [11-17] is incorrect; however, it is not implied correctly. Matter waves can be considered analogous to light waves, but a body cannot be regarded analogous to photons. De Broglie considered the body, in itself, as the quanta of associated matter waves [11-17]; analogous to photons. For this reason, the quantum energy of matter waves ($h\nu$) equals the total energy ($\gamma mc^2$) of body, is proposed by him [11-17]. Consideration of a body to be analogous to a photon is inappropriate, *because they differ in many aspects which are discussed in detail in the upcoming sections.* Hence, the body in itself cannot be regarded as the corpuscle of matter waves, analogous to photons.

The matter wave's frequency expression is inadequate due to the inappropriate analogy implementation between light waves and matter waves. *The rectifications in the concepts and refinement in the analogy implementation between light waves and matter waves are presented. The corpuscular concept introduced for matter waves treats the corpuscle of matter wave as a separate entity, though associated with the body, in a moving mass system.* The concept thus provides a modified frequency and phase velocity expression for matter waves. It is shown that the *frequency of matter waves should necessarily be related to the kinetic energy of the body,* instead of the total energy. Thereafter, a modified expression for the phase velocity of matter waves is also proposed. *The presented rectified expressions for the frequency and phase velocity of matter waves are free from all the inadequacies and negations,* confronted by de Broglie's proposed expressions; for instance, the **modified frequency** expression becomes zero when the body is at rest, supported by the infinite wavelength obtained from $\lambda=h/\gamma m\upsilon$.

The experimentally verified relation $\lambda=h/P$ [30-36] is also derived satisfactorily by means of the proposed modified frequency expression. Further, in section 12, the *dissimilarity among matter and mass is presented,* which actually clarifies the difference between the body and the corpuscle of the associated matter waves. The concept is additionally employed to explain a few theoretically posed questions of relativity, such as mass increment and charge invariance of a moving body. Though both radiation and matter possesses dual character, the reason for this dualism is not described in the literature to the best of our knowledge. Section 15 elucidates the wave-particle duality picture on the basis of the proposed concepts. The result emanating suggests that the particle type characteristics should be possessed by all the waves. Even in the case of matter, particle type character is an effect of the associated matter waves and its corpuscle, and is independent of the presence of matter. Therefore, this concept provides a *justification for the possession of particle characteristics by light, even in the absence of matter. Dual characteristics are revealed as a feature of every wave, thereby providing enormous new experimental opportunities for its verifications.*

The modifications introduced in the frequency expression for matter waves results to the variation in the wave function expression for a freely moving particle and its energy operators with appropriate justifications provided in the paper. **The satisfactory derivations of fundamentally correct Schrödinger's equations, Dirac's equation and Klein-Gordon's Equations** by utilizing the modified expression for the wave function of a freely moving particle and its energy operators provides **a primary verification step for the modifications.**

**Relation between the Kinetic energy and Momentum of the moving body** is also proposed and is subsequently applied to introduce **General and Relativistic Quantum Mechanical Wave Equations**. The transformation of these equations into the Schrödinger's and Dirac's equations is also presented, which provides another verification step for the proposed theory. **Applications of these equations in some bound state quantum mechanical systems, such as Quantum harmonic oscillator, Rigid rotor etc. are presented in the paper.** These equations provide the information regarding particle's general behavior in such systems, valid for both non-relativistic and relativistic cases. The separate application of relativistic wave equation provides the information about particle's behavior, if moving at relativistic velocity. Obtained results can be well transformed into the non-relativistic results, thus providing **another confirmation step for the results**. These equations can further be employed for the study of Relativistic Quantum Mechanics because they provide information of the particle's behavior in trouble-free and straightforward way, analogous to Schrödinger's equations.

Apart from all above optimistic points, **a mathematical proof for the proposed modified frequency and phase velocity expression is also presented.** Two expressions, which are experimentally and conceptually correct, proposed by de Broglie, are revealed to contradict his own proposed frequency expression. Moreover, the expression thus obtained satisfies the modified frequency expression for matter waves, proposed in the paper. **This provides significant evidence, in favor of the proposed modified frequency and phase velocity expression for matter waves.**

Section 28 provides the proof regarding the proposed phase velocity expression. The first part of the section presents the conversion of Schrödinger's equation into the universal wave equation given as:-

$$\frac{\partial^2\Psi}{\partial x^2}=\frac{1}{\upsilon_P^2}\frac{\partial^2\Psi}{\partial t^2}$$

Since this form of wave equation involves the phase velocity term, therefore, the substitution of correct $\upsilon_P$ expression should yield the time dependent Schrödinger's equation. In the second part of the section, **it is shown that the substitution of de Broglie's proposed phase velocity expression ($c^2/v$) fails to yield the correct result; however, the proposed modified phase velocity expression succeeds in the task**. Therefore, the process becomes as a proof regarding the correctness of the proposed phase velocity expression.

Nowadays, almost after 87 years from de Broglie's proposed concepts of matter waves; we have entered an era of matter wave amplification [41-44] by Bose-Einstein condensate. Therefore, it is of vital significance to uncover and rectify the hidden inadequacies associated with the **wave particle duality and matter wave theory, proposed by Louis de Broglie.**

## 2. DE BROGLIE'S PROPOSED CONCEPTS AND ASSOCIATED PROBLEMS

In this section, we have presented the conceptual inadequacies, in the theory proposed by de Broglie in his dissertation [14], research papers [11-13,15,17] and the explanation given by him during his Nobel lecture [16].

The chief ideas which led de Broglie to hypothesize the wave nature of matters [16] are:-

The energy expression, $E=hv$, for the corpuscle of radiation, is unsatisfactory for the corpuscular theory, since it contains the frequency term. Similarly, the concept of atomic motion quantization of Bohr's theory is similar to the situation of interference and Eigen vibrations because they involve whole numbers. The two ideas are a combination of wave and particle concepts. Therefore, he assumed that wave particle characteristic should be exhibited by both radiation and matter.

The concepts followed by de Broglie for the development of theory are presented in this section and the inadequacies associated with these concepts are mentioned.

### 2.1 Concept-1

The fundamental idea pertaining to quanta (body) taken up by de Broglie was the impossibility to have the corpuscle's energy without associating a frequency to it. Therefore, he assumed [11-13,14] that every body of inertial mass $m$ is related to some internal periodic phenomenon of frequency $\nu_0$, measured from the rest frame of the body, such that:-

$$h\nu_0 = mc^2 \tag{1}$$

#### 2.1.1 Inadequacy in concept-1

Expression (1) is a result of the analogy taken up by de Broglie between photons and the body. The energy of a radiation corpuscle is $h$ times the frequency of light wave. Similarly, de Broglie considered body in itself as a quantum/corpuscle, **analogous to photon**. Hence, the energy of body is also assumed to be related to the frequency of the internal periodic phenomenon in a similar manner. However, *a body should not be considered analogous to matter less photon.*

It seems that as photons possess particle type nature; body has intrinsic particle type characteristics, and thus they are analogous. Therefore, the energy of the body is similar to the energy of the photon. However, the question which arises here is: Does body (matter) actually possess particle type nature in itself? Light is matter less and still possesses particle nature!!!! The possession of particle characteristics by matter less radiation put a question on the true identity of a 'particle'. *Whether electrons (matter) are true particles or matter less photons?* The explanation of this anomalous concept of particle type character is explained in section-15.

### 2.2 Concept-2

After assuming the presence of an internal periodic phenomenon having frequency $\nu_0$, associated with the body, de Broglie continued as follows [11-13,14]:

According to the Lorentz transformation for time, the periodic phenomenon in a moving body, as viewed by the fixed observer, is slowed down by a factor of $\sqrt{1-\beta^2}$ , where $\beta=\upsilon/c$ and $\upsilon$ is the velocity of moving body. Hence, the frequency of the periodic phenomenon, for the fixed observer, is manifested according to relativity concepts, as follows:-

$$\nu' = \nu_0\sqrt{1-\beta^2} \tag{2}$$

where $\nu'$ is the frequency of the phenomenon within the body, viewed from the reference frame of the fixed observer which is in relative motion with respect to the body. Substituting equation (1) in (2):

$$\nu' = \frac{mc^2}{h}\sqrt{1-\beta^2} = \frac{mc^2}{h\gamma} \tag{3}$$

This frequency expression is obtained by the Lorentz transformation process for the internal periodic phenomenon of the body.

On other hand, from de Broglie's own concept [11-17], also mentioned previously, the energy of a moving object ($\gamma mc^2$) should be related to the frequency, according to the quantum energy relation ($E=h\nu$), which gives frequency to be:-

$$h\nu''= mc^2 \frac{1}{\sqrt{1-\beta^2}} = \gamma mc^2 \qquad (4)$$

$$\Rightarrow \nu'' = \frac{mc^2}{h}\frac{1}{\sqrt{1-\beta^2}} \qquad (5)$$

where $\nu''$ is the frequency of the internal periodic phenomenon of the body, viewed from the reference frame of the observer, but this time it is obtained from the concept presumed by de Broglie.

Hence, two different frequency expressions [eq. (5) and (3)] are obtained, by de Broglie. One is obtained by employing the Lorentz Time Transformation [equation (3)]; whereas other is due to the consideration of body in itself as the quantum of matter waves, of energy $h\nu$, by de Broglie.

To explain these two different frequency expressions, de Broglie gave the 'theorem of phase harmony' stating [14]:-

*'A periodic phenomenon is seen by a stationary observer to exhibit the frequency* $\frac{mc^2}{h}\sqrt{1-\beta^2}$ *that appears constantly in phase with a wave having frequency* $\frac{mc^2}{h}\frac{1}{\sqrt{1-\beta^2}}$ *propagating in the same direction with velocity* $c/\beta$ *.'*

Here de Broglie considered that the frequency expression ($\nu'$) obtained by Lorentz transformation process, presents the frequency of internal periodic phenomenon, viewed from the frame of reference of the fixed observer. Whereas, the frequency $\nu''$ is the frequency of some wave associated with the moving body; which travels at a velocity $c/\beta$, to appear always in phase with the phenomenon.

### 2.2.1 Inadequacy in concept-2

To explain the two dissimilar frequency expressions, de Broglie considered that the frequency $\nu''$ is the frequency of some associated wave and *not of internal periodic phenomenon of the body.* The initial consideration for the existence of internal periodic phenomenon of body and then its segregation in terms of a wave is misleading. Both the frequency expressions [(3) and (5)] should represent the internal periodic phenomenon of body, viewed from a frame of the fixed observer.

Moreover the frequency $\nu''$, in previous section, is again the result of analogy taken up by de Broglie, between body and photons, i.e. body in itself as the corpuscle, *analogous to photon*, possessing energy $h\nu$. Therefore, the term $h\nu$ is substituted equivalent to the energy of body, $\gamma mc^2$.

## 2.3 Concept-3

Subsequent to the consideration of two different periodic oscillations (wave and phenomenon), de Broglie proved them to be in phase, i.e. his proposed phase harmony concept. De Broglie proposed that [17,14], if the phenomenon and wave is supposed to maintain phase harmony at *t=0*. Then after time *t,* the moving object traverses a distance $x=\beta ct$, Therefore, phase of the periodic phenomenon of the body develops into:-

$$\nu' t = h^{-1} mc^2 \sqrt{1-\beta^2}\,(x/\beta c)$$

However, to represent the phase of the *wave traversing the same distance,* de Broglie introduced a time lag ($\tau$) of $\beta x/c$, [14] i.e.:-

$$\nu''(t-\tau) = \frac{mc^2}{h}\frac{1}{\sqrt{1-\beta^2}}\left(\frac{x}{\beta c} - \frac{\beta x}{c}\right) = \frac{mc^2}{h}\frac{x}{\beta c}\sqrt{1-\beta^2}$$

In this fashion, the phase equality among body's phenomenon and its associated wave is demonstrated by de Broglie.

Consequently, the frequency and phase velocity expression presented by de Broglie are:-

$$\upsilon_p = \frac{c}{\beta} = \frac{c^2}{\upsilon} \text{ and } h\nu = \frac{mc^2}{\sqrt{1-\beta^2}}$$

### 2.3.1 First inadequacy in concept-3

The postulation of identical phase matching (phase harmony) condition, at $t$=0, between the phenomenon and wave, implies that there is no phase or time difference between them because both of them begin simultaneously. However, after a certain time $t$, the consideration of a finite time difference ($\tau=\beta x/c$) for wave, stand out against the phase match condition, assumed by de Broglie. The time difference honestly implies that wave initiate its propagation after a time of $\beta x/c$, from the initiation of body's journey at $t=0$. This further implies that body and wave would remain separated for that particular time interval of $\beta x/c$. This diverges from de Broglie's proposed concept of associating the moving body with a matter wave, continuously during its motion.

Moreover, he considered that *the time difference introduced, $\beta x/c$, is equal to the time taken by wave itself, to travel the distance x;* since the velocity of wave assumed by de Broglie is $c/\beta$. Hence, the whole concept of time difference is utilized in an inappropriate way.

### 2.3.2 Second inadequacy in concept-3

The phase velocity of the matter waves ($c/\beta$) was *assumed* by de Broglie, to show the existence of phase harmony condition. However, the phase harmony is itself not correct as mentioned in the previously. Therefore, the phase velocity of matter waves could not be equal to $c/\beta$.

The velocity of matter waves proposed by de Broglie is always found to be greater than the speed of light in vacuum and opposes the principle of Special Theory of Relativity.

## 2.4. Concept-4

De Broglie presented another derivation for the frequency and phase velocity relation [14,16] as follows:-

To establish a certain parallelism between the motion of corpuscle (body) and the propagation of the associated wave, we consider a reference system ($x,y,z,t_0$). In this system, the corpuscle is immobile and forms the intrinsic Galilean system. De Broglie stated that [16]:-

In the rest frame of a body, *the wave should be stationary since the corpuscle is immobile; its phase is same at every point.* The wave, in this frame, is represented as:

$$\sin 2\pi\nu_0(t_0) \qquad (6)$$

where, $\nu_0$ is the frequency of the wave in the rest frame of corpuscle (body) and $t_0$ is the instantaneous time of this frame.

### 2.4.1 Inadequacy in concept-4

In compliance with de Broglie's concept, the phase of matter wave should be same everywhere in the intrinsic frame of the body; i.e. it should be a constant for the zero velocity of body. However, substitution of eq. (1) (frequency of matter wave when body is at rest) in eq. (6), gives:-

$$\sin 2\pi\frac{mc^2}{h}(t_0) \qquad (7)$$

Since, the phase in wave equation (7) involves a time variable $t_0$, it is not a constant. **This is a self contradiction to de Broglie's concept,** stated above. The contradiction **evidently** indicates that there is some problem in the frequency expression proposed by him, because it cannot produce a constant phase of matter wave, associated with the body at rest. **The concept of constant phase proposed by de Broglie is conceptually correct**, in view of the fact that matter waves originates due to body's motion. Consequently, a Galilean frame in which body is immobile, the associated matter waves should also become stationary. However, the frequency expression proposed by him is inappropriate; therefore, the mathematical harmony of the concept does not meet.

## 2.5 Concept-5

To proceed further, de Broglie considered [14,16] another frame ($x',y',z',t$), in which the corpuscle (body) is moving with a velocity $v=\beta c$, with respect to the observer. In compliance with Lorentz transformation, time $t$ taken by the observer will be associated with the intrinsic time of the corpuscle $t_0$ *as*:

$$t_0 = \frac{t - \frac{\beta x}{c}}{\sqrt{1-\beta^2}}$$

The above expression is substituted in the wave equation (6), to give:-

$$\sin 2\pi \frac{\nu_0}{\sqrt{1-\beta^2}}\left(t - \frac{\beta x}{c}\right) \qquad (8)$$

Comparison of equation (8) with the general equation of wave:-

$$A \sin 2\pi \nu'' (t - \tau) = A \sin 2\pi \nu'' (t - \frac{x}{\upsilon_P})$$

yielded the frequency and phase velocity of the associated wave to be:

$$\nu'' = \nu_0 \Big/ \sqrt{1-\beta^2} \qquad (9)$$

$$\upsilon_p = \frac{c}{\beta} = \frac{c^2}{\upsilon} \qquad (10)$$

De Broglie argued [11-17] that in the intrinsic Galilean reference system of the corpuscle, the energy of
corpuscles equivalent only to its rest energy ($mc^2$),i.e.

$$h\nu_0 = mc^2 \qquad (11)$$

Substitution of= $mc^2/h$ in Eq. (9), provided:-

$$\nu'' = \frac{mc^2}{h}\frac{1}{\sqrt{1-\beta^2}}$$

$$\Rightarrow h\nu'' = \frac{mc^2}{\sqrt{1-\beta^2}}$$

The frequency expression thus obtained is identical to equation (5). Therefore, the frequency of matter waves is proposed by de Broglie, to be related to the total energy of body, i.e. for matter waves [11-17]:-

$$\nu = \frac{\gamma mc^2}{h} \qquad (12)$$

### 2.5.1. First inadequacy in concept-5

The deeper analysis of concept-5 reveals that de Broglie considered the body in itself to be the corpuscle of the associated matter wave. Equation (11) is also the result of this consideration only. It implies that the body is analogous to photons (since light wave is kept analogous to matter waves). Elaborately, the analogy taken by de Broglie is:-

If the frequency of light wave is *f*, then the corpuscle of light (photons) carries energy *hf*. Similarly, if the body is also associated with some kind of waves of frequency $\nu_0$, then the corpuscle of this wave also possesses energy, $h\nu_0$ (in analogy with photons).The energy of body at rest is equal to its rest energy, $mc^2$.

*De Broglie considered body in itself to be the corpuscle of the associated wave, and hence the energy of the body ($mc^2$) equals the energy of corpuscle ($h\nu_0$)*, as in Eq. (11).

The last point of analogy, that the body is analogous to photon, is not correct because photons and body (matter) differ from each other in terms of inertial energy. *Photons do not have inertial mass, whereas a body will possess it, for sure. Hence, they are not analogous.*

### 2.5.2 Second inadequacy in concept-5

The term $\beta x/c$ involved in the Lorentz time transformed equation (8), is essentially treated by de Broglie as the *initial time (time difference)* of the wave, and thus/$\beta$ became the *superluminal* phase velocity of matter wave accordingly. However, it is not acceptable. The term $\beta x/c$ occurs only due to Lorentz time transformation formula. It is not correlated with the initial time (time difference) concept for waves. Consequently, the ascription of matter wave's phase velocity equal to $c/\beta$ is inappropriate.

For a while, if the process of phase transformation for matter wave, followed by de Broglie is assumed to be correct, then the same can also be considered for any other sinusoidal wave. So, the phase velocity of any sinusoidal wave should also be invariantly $c/\beta$. However, this is again not true. Hence, the phase velocity of matter wave cannot be equal to $c/\beta$.

### 2.5.3 Third inadequacy in concept-5

In the Lorentz Time Transformation of wave equation (6), de Broglie transformed the time variable $t_0$ in terms of $t$; however, the same is not followed for the frequency term $\nu_0$ (which is also a function of $t_0$). Similar to the time variable ($t_0$) transformation, frequency transformation should also be considered. The frequency term (because of time dilation) becomes:-

$$\nu_0 = \nu'' \sqrt{1-\beta^2}$$

*This variation in the frequency term is not considered by de Broglie, in the equation transformation process.*

Furthermore, the frequency and phase velocity of matter waves is presented by de Broglie by utilizing equation (8). Equation (8) is obtained by the Lorentz time transformation of wave equation (6), i.e. equation (8):-

$$\sin 2\pi \frac{\nu_0}{\sqrt{1-\beta^2}}\left(t - \frac{\beta x}{c}\right)$$

From the very first look, at this equation it can be deduced that a single wave equation, from the frame of reference of observer, is associated with two time variables $t$ and $t_0$ ($t_0$in terms of $\nu_0$). However, this is definitely not possible. A wave equation should be guided only by a single time variable.

### 2.6. Concept-6

By the assistance of the frequency ($h\nu=\gamma mc^2$) and phase velocity expression ($c^2/\upsilon$), wavelength expression for the matter waves associated with a moving body is estimated by de Broglie [11-17]. The phase velocity ($\upsilon_P$') of any wave is related to its wavelength and frequency as [11-17]:-

$$\upsilon_P' = \lambda \nu \tag{13}$$

Substitution of equation (10) and equation (12) in eq. (13) provides:-

$$\lambda = \frac{c^2}{\upsilon}\frac{h}{\gamma mc^2} = \frac{h}{\gamma m\upsilon} = \frac{h}{P} \tag{14}$$

Equation (14) relates the momentum of body and the wavelength of the associated matter wave [11-17]. It is one of the triumphs of de Broglie's theory and has many experimental confirmations [30-36].

### 2.7. Concept-7

In this section, the matter wave's group velocity concept, proposed by de Broglie [14-16] is discussed and is entirely accepted.

Equation (10) dictates the phase velocity of matter waves to be always greater than the speed of light in vacuum. For the explanation of this undesirable result, de Broglie proposed the existence of matter waves in fashion of a group wave, instead of a pure monochromatic phase wave. The calculation of group velocity ($\upsilon_g$) of matter wave is found to be equal to velocity of body ($\upsilon$) [14-16], i.e.

$$\upsilon_g = \upsilon$$

Uncertainty Principle, $\Delta E \Delta t \geq h/4\pi$ [26-27], prohibits the existence of a pure monochromatic wave. For this reason, the matter waves associated with the body should also be a group wave. Consequently, the inclusion of group velocity concept, proposed by de Broglie, is entirely acceptable. However, we should mention here that though the group velocity concept and expression is acceptable, the phase velocity expression is still inadequate and non-acceptable (due to the inadequacies associated with it).

### 2.7.1. Inadequacy in phase harmony concept due to the appropriateness of concept-7

The group wave existence of matter waves, opposes the phase harmony {section 2.2 and 2.3} concept among wave and periodic phenomenon of body, proposed by de Broglie. The physical existence of matter wave in fashion of a group wave, implies its propagation at the velocity equivalent to the group velocity ($\upsilon_g=\upsilon$), instead of the phase velocity ($\upsilon_P=c/\beta$). However, *from de Broglie's proposed phase harmony concept, presented in section 2.2 and 2.3, the propagation of matter wave at the phase velocity c/β is an essential requirement for the satisfaction of continual phase matching condition among the periodic phenomenon of body and the associated matter wave.* However, this is surely not achievable for the reason that the matter waves should propagate at the group velocity, $\upsilon_g=\upsilon$. This opposes the accomplishment of phase harmony among periodic phenomenon and the associated matter wave. Therefore, the superluminal phase velocity expression *c/β* and phase harmony concept proposed by de Broglie, becomes inappropriate simultaneously.

## 3. WAVELENGTH OF MATTER WAVES ASSOCIATED WITH THE BODY AT REST

*This section supports the concept of matter wave's infinite wavelength associated with the body at rest, proposed by de Broglie.* By means of this valid concept, numerous inadequacies of the frequency and phase velocity expression are brought up in the upcoming sections.

The concept of maintaining a constant phase of matter wave in the rest frame of body, proposed by de Broglie, presented in section 2.4, is conceptually correct; since matter waves originates due to the motion of the body. Therefore, a Galilean frame in which the body is at rest, the associated matter wave should also becomes stationary, as represented in figure 1.

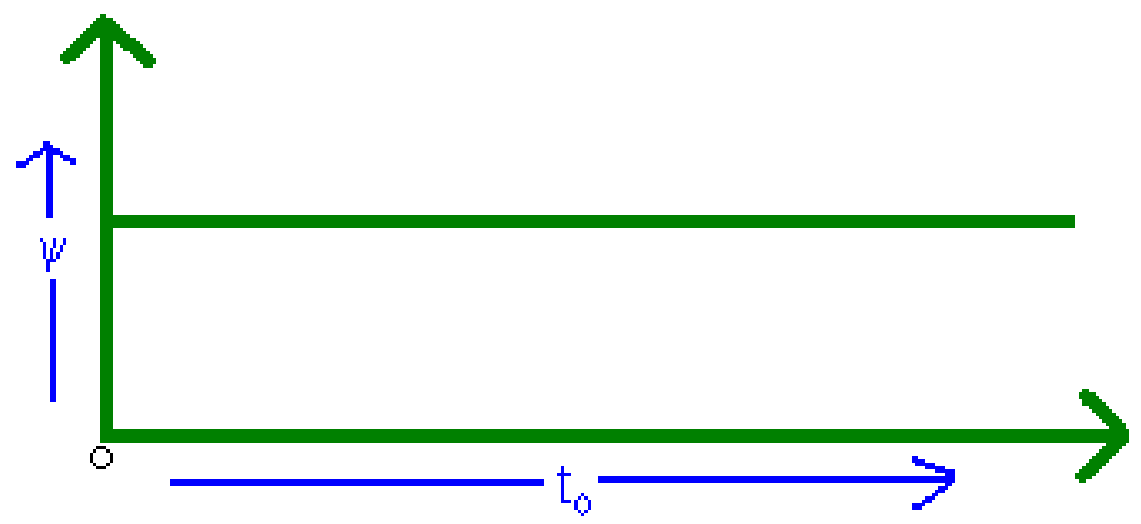

**Figure 1.Representation of matter wave in rest frame of the body.**

Figure 1 illustrates the stationary matter waves maintaining a constant phase, associated with the body at rest. The infinite (∞) wavelength of the matter wave associated with the body at rest is admissible by means of figure (1).

Moreover, the experimentally verified wavelength expression [30-36], proposed by de Broglie, $\lambda=h/P$, also favors the infinite wavelength of matter waves corresponding to the body at rest. *Therefore, the wavelength of matter wave tends to infinity, when the velocity of body is zero.*

## 4. MATHEMATICAL AND CONCEPTUAL INCOMPATIBILITIES OF MATTER WAVE'S PHASE VELOCITY EXPRESSION

The additional incompatibilities of the phase velocity expression, proposed by de Broglie, are discussed in this section.

(1) The phase velocity ($\upsilon_p$) expression for matter waves given by de Broglie is [eq. (10)]:-

$$\upsilon_p = c^2/\upsilon$$

The phase velocity of matter waves is always greater than *c*, which opposes the principle of special theory of relativity [5,45-52].

(2) De Broglie referred the phase velocity of matter waves as insignificant in his dissertation [14], by proposing their energy less existence. However, this is not possible because, in that situation, the group wave, which is the superposition of every smaller phase waves, would also become energy less and the wave would not be able to propagate. Moreover, equation [(1), (5), (11), and (12)] reveals that the waves should possess energy, for sure, even from de Broglie's view.

(3) The phase velocity of matter wave corresponding to the body at rest from eq. (10) is infinite (1/0, ∞). This value of phase velocity has no solution. Simultaneously, from section 3, the wavelength of the matter wave, when body is at rest, is infinite (1/0) and the frequency corresponding to an infinite wavelength should be zero. So, the phase velocity from the expression $\upsilon_p = \lambda\nu$, [53-55] would certainly become 0/0, which has many solutions.

HENCE, the two concepts are mathematically in contrast with each other. One ($\upsilon_P=c^2/\upsilon$) gives no solution, whereas the other ($\upsilon_P=\nu\lambda$) gives many solutions. *Since ($\upsilon_P=\nu\lambda$) could not be falsified, therefore ($\upsilon_P=c^2/\upsilon$) should be the reason for this inadequacy.*

(4) Incorrectness in the phase velocity expression, ($c^2/\upsilon$), can also be analyzed as:

If a body is at rest, the individual phase waves possess infinite velocity ($\upsilon_P=c^2/0$), whereas the group wave has zero velocity, since $\upsilon_g=\upsilon$ (from de Broglie's concepts).

(5) The phase velocity expression follows an inverse relation with the velocity of body ($\upsilon$), whereas the group velocity is directly proportional to it. If velocity of body increases, then its phase velocity would consequently decrease; whereas the group velocity increases.

*Consequently, in this section, we found that the phase velocity expression given by de Broglie is inadequate.*

## 5. MATHEMATICAL AND CONCEPTUAL INCOMPATIBILITIES OF MATTER WAVE'S FREQUENCY EXPRESSION

Numerous additional evidences validating the inadequacy persisting in the frequency expression are discussed in this section.

**First**, from Equation (12):-

$$h\nu = \gamma mc^2$$

If the body is at rest, then:-
From section 3, the wavelength of matter waves becomes $\infty$. Therefore, the frequency should become zero. Hence, the L.H.S. term of the frequency expression becomes zero and R.H.S becomes $mc^2$

⇨ L.H.S. ≠ R.H.S as $0 \neq mc^2$.

Therefore, the frequency expression is mathematically inadequate.

**Second**, the infinite wavelength of matter waves, justified in section3, corresponding to the zero velocity of body, represents that the distance between two consecutive planes of same phase of wave is infinite. Or, the wave has to traverse an infinite distance to appear in same phase again. Consequently, the time period of the wave also becomes infinite. Conversely, *the frequency of matter wave corresponding to the body at rest should invariably become zero.* Conceptually also, the frequency corresponding to the infinite wavelength should be zero. However, *the matter wave's frequency expression (12), proposed by de Broglie, dictates a finite non-zero value of frequency equal to $mc^2/h$.* This demonstrates the inappropriateness of the frequency expression.

**The finite frequency thus obtained is in contrast to the infinite wavelength of matter waves (when body is at rest)**. Since the wavelength expression ($\lambda=h/P$) is an experimentally verified relation [30-36], it could not be challenged. This evidently implies that the frequency expression must be held responsible for this inadequacy.

**Third**, from '*concept-4*' of section 2, the waves should be stationary when the body is at rest. However, the finite value of frequency $\left(mc^2/h\right)$ opposes this concept. Finite frequency value of the wave is possessed only in course of its propagation. Hence, neither the wave, nor the body should be at rest which is contradicting the fact that body considered in this situation is at rest.

*Consequently, it is now clear that the frequency expression given by de Broglie is inadequate.*

## 6. RECTIFICATIONS IN CONCEPTS

In earlier sections, the inadequacies persisting in the theory and concepts of matter waves, proposed by de Broglie, are presented. The inappropriate concepts lead to inadequate frequency and phase velocity expression for matter waves. The inadequacies in the concepts are also mentioned at the same time. However, **some conceptually correct concepts, proposed by him {section 2.4 and 2.7}, are supported wholeheartedly**. Therefore, the concepts proposed by de Broglie are only partially correct. These inappropriate concepts actually led to an inadequate expression for frequency and phase velocity for matter waves. Actually the inadequacy in de Broglie's proposed theory is the result of his inappropriate analogy among light waves and matter wave {section 2.1.1, 2.2.1 and 2.5.1} i.e. photon and body are treated to be analogous. It was considered that as photon is responsible for the particle characteristic of light, body (matter) in itself possesses intrinsic particle type nature. Hence, body on its own is responsible for its particle type behavior and thus analogous to photon. Therefore, due to the analogy between photon and body, the energy ($\gamma mc^2$) of body should be related to the frequency of the associated matter waves according to the relation that holds for photons, i.e.:-

$$h\nu = E = \gamma mc^2$$

However, the concept is not appropriate. Photons can never be made analogous to body (matter), because they differ from each other in terms of inertial energy {section 2.5.1}.

Actually the possession of particle characteristic by radiation puts up a question on the possession of particle characteristic by matter. *How can a matter less radiation possess those characteristics which are observed only for matters? Therefore, the question regarding the true identity of a 'particle' arises. Whether matter, like $e^-$, is truly a particle or photons?* This delicate concept was overlooked and thus it resulted to the inadequate matter wave's concepts and expressions (frequency and phase velocity).

In this section the rectifications in the concepts of matter waves and its related theory are presented as follows:

If a wave (light) possesses particle type character, then the body (matter) should also have some kind of waves (matter waves) associated with it. These waves (matter waves) should originate due to the motion of body. Therefore, a frame in which the body is at rest, the associated matter waves should also become stationary. Consequently, the phase of the waves should become constant in that particular frame of reference. This further implies that the frequency and wavelength of matter wave should be zero and infinite respectively, when body is immobile.

*Analogy between light waves and matter waves provides:* If the frequency of light wave is $f$, then the corpuscle of light possesses energy $hf$. Similarly, if a moving body is also associated with matter waves of frequency $\nu$, then its corpuscle should possess energy $h\nu$, *analogous to light waves.*

**For the analogy among matter waves and light waves, to hold correct; the existence of matter wave's corpuscle (analogous to photons) is necessary. However, the body (matter) in itself is not the corpuscle of the associated matter waves; otherwise body would become analogous to photon, which is inappropriate. As the light wave is associated with photons; similarly matter waves should also be accompanied with some sort of corpuscle, *but the body in itself is not the corpuscle of matter waves*. Here, the corpuscle is considered as an entity, dissimilar and distinct from the body itself. However, it is always associated with the body, due to the continual correlation of matter waves with the body. *This corpuscle is actually analogous to photons, instead of the body itself.***

Since we have not considered body in itself as the quanta of matter waves, we do not apply the Internal Periodic Phenomenon concept to it, as taken by de Broglie {section 2.1, 2.2}.
The energy of matter wave's corpuscle is related to the frequency of associated matter wave by the relation similar that holds for photons, i.e.:-

*Energy of matter wave's corpuscle=hv*

Since the body in itself cannot be the corpuscle of matter wave; hence the energy of corpuscle of matter waves is not equal to the energy of body, i.e.:-

$$h\nu \neq \gamma mc^2 \neq \frac{mc^2}{\sqrt{1-(\upsilon/c)^2}}$$

Corrected analogy implementation among light waves and matter waves provides a decisive solution for the energy expression of matter wave's corpuscle, in the upcoming section.

## 7. CORRECTED IMPLEMENTATION OF ANALOGY BETWEEN LIGHT WAVES AND MATTER WAVES

In terms of energy, photons possess only the $hf$, where $f$ is the frequency of light wave. Since photons are mass less, therefore, $hf$ is their total energy. Thus, $hf = \gamma mc^2$ is *mathematically* true for photons. Here, it seems that for the correctness of analogy among photon and matter wave's corpuscle, the corpuscle of matter waves should also possess energy equal to the total energy of the moving mass system. However, this is not true.

Explanation:-

For photons, mathematically, *hf* is the total energy ($hf = \gamma mc^2$ = Kinetic Energy + Rest Energy), *however, conceptually it is only their kinetic energy, because of the absence of rest energy* (photons are matter-less).

Simultaneously, in a moving mass system, total energy is actually the sum of its non-zero rest energy and the kinetic energy. Hence, the total energy of body and photon conceptually differs from each other. Therefore, $h\nu = \gamma mc^2$ is true for photons, but not for the corpuscle of matter wave.

Due to the consideration of analogy among matter waves and light waves, the energy of the corpuscle of matter wave cannot be designated to be the total energy of moving-mass system. As discussed in previous section, total energy is possessed by the body itself. Therefore, it leads to the consideration of body in itself to be the corpuscle of the associated matter waves; which further implies a direct analogy between photons and body; and is not true. Hence, corpuscle of the matter wave is a separate entity but is associated with the moving body, because matter waves are also associated with the body.

Combining all these concepts, it is found that **the analogy of matter waves and light waves holds correct, only if the energy of photon and corpuscle of matter wave are similar**. For photons, *hv* is conceptually their kinetic energy **hence it should be only kinetic energy for the case of matter wave's corpuscle also, i.e. for matter waves**:-

$$
\begin{aligned}
h\nu &= \textit{kinetic energy of system} \\
&= \textit{(total energy)} - \textit{(rest energy)} \\
\Rightarrow h\nu &= \gamma mc^2 - mc^2 = (\gamma - 1)mc^2 \qquad (15)
\end{aligned}
$$

**Consequently, the proposed modified frequency expression of the matter wave associated with a moving body becomes**:-

$$
\nu = \frac{K}{h} = \frac{mc^2}{h}\left(\frac{1}{\sqrt{1-\left(\frac{\upsilon}{c}\right)^2}} - 1\right) \qquad (16)
$$

where, $K$ is the kinetic energy of the moving mass system.

The modified frequency expression for matter waves also supports the following concepts:-

**First,** matter wave originates due to the motion of the body. Therefore, matter wave should possess energy only when the body is mobile. By virtue of motion, body possesses the Kinetic energy. So, the energy of the matter wave's corpuscle should be equivalent to the kinetic energy of the system.

**Second**, since the body in itself cannot be the corpuscle of the associated matter waves. So, the sum of the energy of the body (inertial), $mc^2$ and the energy of corpuscle of matter wave, $h\nu$, should contribute to the total energy of the system, $\gamma mc^2$ i.e.

$$
\begin{aligned}
h\nu + mc^2 &= \gamma mc^2 \\
\Rightarrow h\nu &= (\gamma - 1)mc^2
\end{aligned}
$$

A noticeable point here is that, if a body moves with the speed almost equal to that of light in vacuum (*c*), the contribution of rest energy to the total energy of the system becomes negligible. In that situation, the total energy of moving body mainly comprises of the kinetic energy and thus *hv* becomes nearly equal to $\gamma mc^2$. **Therefore, de Broglie's proposed frequency expression is valid only when the velocity of body becomes almost equal to the speed of light in vacuum (*c*).** This could be true only when velocity of body becomes almost equal to *c*.

## 8. GROUP VELOCITY OF MATTER WAVES

In this section, we have shown that the matter waves associated with a moving body is a group wave, having velocity equal to the velocity of the body; *on the basis of the fact that the energy associated with* *the matter waves is only the kinetic energy of the system.*

Uncertainty principle ($\Delta E \Delta t \geq \hbar/2$) dictates the non-existence of a pure monochromatic phase wave [27,28]. For this chief reason, the matter wave associated with the body should also be a group wave. Group velocity of any wave refers to the velocity with which the energy of the wave is transported [53].Since, the energy associated with matter wave is only the kinetic energy of the moving mass system; therefore, the group velocity of matter wave should be equal to the velocity of body:-

$$\upsilon_g = \upsilon \qquad (17)$$

Group velocity equal to the velocity of body is directly postulated, on the basis of the energy associated with matter waves. Group velocity is the actual velocity of matter wave and is of prime physical significance.

## 9. DERIVATION OF THE WAVELENGTH EXPRESSION (λ=*h/P*)

In this section we have derived the experimentally verified relation [30-36] $\lambda = h/P$, by employing the proposed modified frequency and phase velocity expression. Group velocity of waves is given by [28,29,53,54]:

$$\upsilon_g = \frac{d\nu}{d\left(\frac{1}{\lambda}\right)}$$

Or,

$$d\left(\frac{1}{\lambda}\right) = \frac{d\nu}{\upsilon_g} \qquad (18)$$

Substituting equation (16) and (17) in equation (18):-

$$d\left(\frac{1}{\lambda}\right) = \left(\frac{m}{h}\right)\left[1-(\upsilon/C)^2\right]^{-3/2} d\upsilon$$

Integrating the above equation using proper limits provides:-

$$\int_0^{\frac{1}{\lambda}} d\left(\frac{1}{\lambda}\right) = \frac{m}{h}\int_0^{\upsilon}\left[1-\left(\frac{\upsilon}{C}\right)^2\right]^{-3/2} d\upsilon$$

$$\Rightarrow \frac{1}{\lambda} = \frac{P}{h} = \frac{\gamma m \upsilon}{h} \qquad (19)$$

Therefore, the experimentally verified [30-36] matter wave's wavelength expression is derived satisfactorily by the proposed modified frequency and group velocity expression. *The satisfactory derivation of the wavelength expression provides the first verification step for the proposed concepts and theory.*

## 10. VERIFICATIONS OF THE MODIFIED FREQUENCY EXPRESSION

In this section, we have shown that the new frequency expression is free from all the inadequacies confronted by de Broglie's proposed frequency expression, discussed in sec. V and II.D.1.

**First**, in section 2.4.1 we discussed the inability of frequency expression for matter wave, proposed by de Broglie, to produce a constant phase, corresponding to the body at rest.

However, the corrected frequency expression (16), when velocity of body is zero, becomes:-

$$\nu_0 = \left(\frac{1}{\sqrt{1-0/c^2}} - 1\right) mc^2 = 0$$

Wave can therefore be represented as:-

$$\sin 2\pi(\nu_0 t_o) = \sin 2\pi(0)t_0 = \sin(0)$$

*The phase of matter wave now becomes zero, which is a constant. So the concept of constant phase of wave, proposed by de Broglie, corresponding to body at rest is in mathematical harmony.*

**Second,** in section 5we have seen that the frequency of matter wave corresponding to the body at rest was contradicting many concepts. However, *the proposed modified frequency expression is in harmony with other concepts also.*

Since $h\nu = (\gamma - 1)mc^2$ is the proposed modified frequency expression. From it, *the matter wave's*

*frequency corresponding to body's zero velocity is zero.* Simultaneously, from Section 3 also, the frequency of matter waves should be zero, because the wavelength becomes infinite. Now, *the two concepts do not contradict each other contrary to de Broglie's proposed frequency expression.*

**Third**, matter wave's zero frequency suggests that the body and its associated waves are at rest which is true. *This is in contrast with discussion given for de Broglie's frequency expression, in section 5.*

*Hence the new frequency expression does not contradict other concepts. Consequently, from this section the appropriateness of the proposed modified frequency expression becomes clear.*

## 11. MODIFIED PHASE VELOCITY EXPRESSION FOR MATTER WAVES AND ITS VERIFICATIONS

Phase velocity of a wave is [28,53-55] related to its frequency and wavelength as:-

$$\upsilon_p = \lambda \nu$$

Substituting equation (16) and (19) in above equation, the phase velocity of matter wave is obtained as:-

$$\upsilon_p = \frac{c^2}{\upsilon}\left[1-\sqrt{1-\left(\frac{\upsilon}{c}\right)^2}\right] \qquad (20)$$

Above equation is the proposed modified phase velocity expression for matter waves. In terms of $\gamma$ it becomes:-

$$\upsilon_p = c\sqrt{\frac{\gamma-1}{\gamma+1}} \qquad (21)$$

However, the phase velocity expression for matter waves proposed by de Broglie is $c^2/\upsilon$. The tailored phase velocity expression (20) and (21), dictates matter wave's phase velocity to be always less than the speed of light in vacuum ($c$), whereas the phase velocity expression proposed by de Broglie is always greater than $c$. For this reason the corrected phase velocity expression does not contradict the Principle of Special Theory of Relativity. Thus, the first inadequacy of section 4 is over.

The term $\sqrt{1-(\upsilon/c)^2}$ of equation (20) can be expanded binomially, if $\upsilon << c$, consequently the phase velocity of matter waves becomes half the velocity of body, i.e. for non-relativistic cases:-

$$\upsilon_p = \frac{\upsilon}{2} \qquad (22)$$

*The obtained phase velocity equation shows if the body moves at non-relativistic speed, then the phase velocity of matter waves is less than the velocity of body itself. However, this does not imply that matter waves are slow and follows the body from behind.* Since, the matter waves associated with the body always exist as group wave [section-8] having *group velocity equal to the velocity of the body* (equation 17). Therefore, matter waves continuously propagate in company of the body. The group velocity expression is of prime physical significance.

The third contradiction confronted by the phase velocity expression, proposed by de Broglie, observed in section 5, is now terminated. The corrected phase velocity expression for matter wave dictates the phase velocity to be0/0, when the body is at rest. Simultaneously, from section 3, the wavelength and frequency of matter waves corresponding to the body at rest is infinite and zero, respectively. Therefore, the relation $\upsilon_p = \nu\lambda$ also provides the same. *The contradiction in this manner is removed.*

The 0/0 value of phase velocity, corresponding to body at rest, can have infinite no. of solutions, with zero also as one of the solution. Moreover, numerically $0 << c$, therefore from equation (22) it can be deduced that the phase velocity of matter wave associated with body at rest is zero. Therefore, matter waves have zero phase and group velocity, when body is at rest. **The fourth inadequacy of section 4 is also removed.** Figure 2 shows the variation of phase velocity of matter waves w.r.t. velocity of body ($\upsilon$), in terms of $\beta = \upsilon/c$.

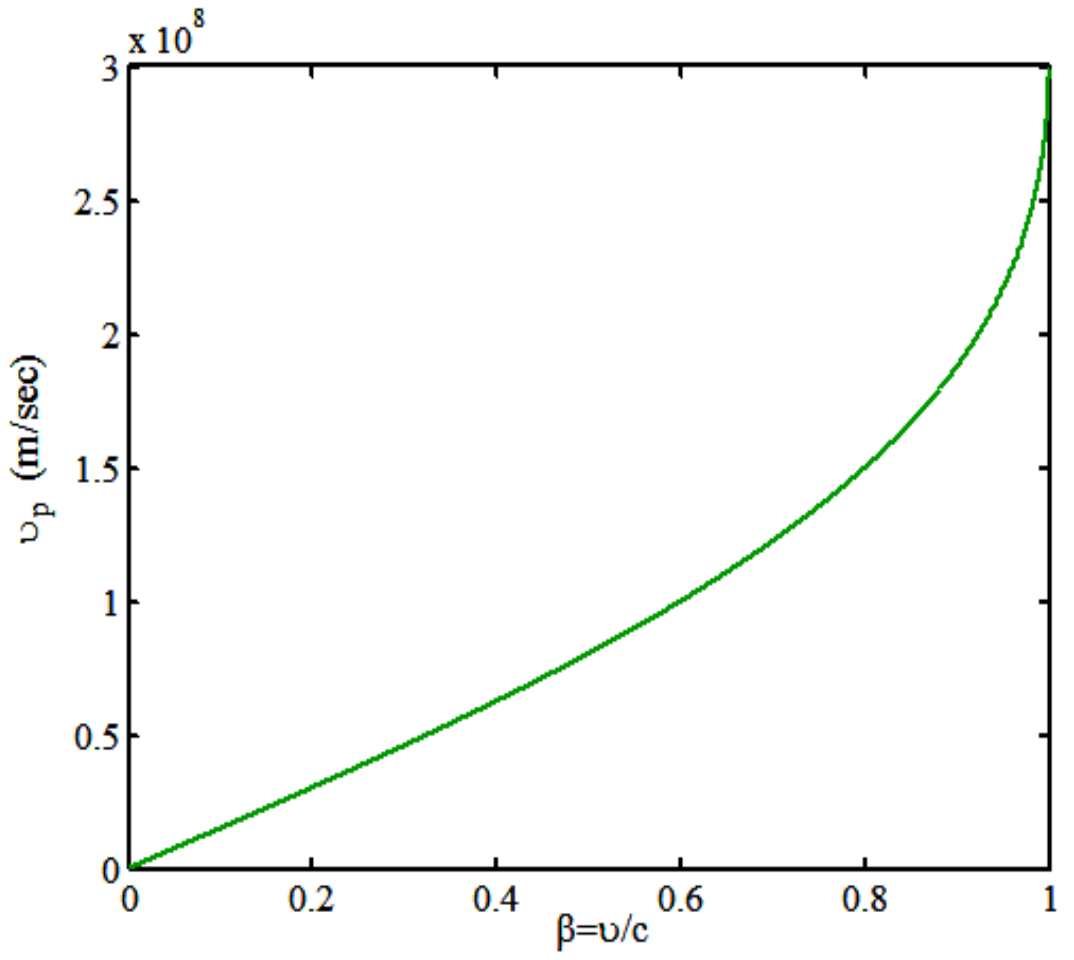

**Figure 2. Variation of phase velocity (modified) of matter wave w.r.t. velocity of body (in terms of *β*)**

For the non-relativistic motion of body, the phase velocity of the associated matter wave is half the velocity of body, i.e. $\upsilon_p = \upsilon/2$.Therefore, curve in figure2, corresponding to the domain of non-relativistic velocity of the body, is a straight line with slope of $\tan^{-1}(c/2)$, whereas the curve corresponding to the domain of relativistic velocity of the body is not linear.

## 12. DISTINCTION BETWEEN CORPUSCLE (MASS) AND BODY (MATTER)

In this section, we have presented the physical interpretation of the corpuscle of matter wave and its distinction from the body itself.

Let us consider the case of a photon. Photons are regarded as mass less particles because from special theory of relativity, a particle traveling at speed equal to speed of light in vacuum (*c*) should have a zero mass [28,47,48,56]. The use of the word *Mass-less*, for photons, is a misnomer. Since photons always possesses energy *hv* [1,4,57-59], *It's mass is therefore,* $hv/c^2$ (Since $E=mc^2$) [5,28,29,45-48].This mass is often coated as the gravitational mass of photons[28], which is in contrast with the zero mass concept for photons provided by relativity.

The contradiction is removed by the deeper analysis of the situation. First, photons carries energy *hv*, hence they possess an equivalent mass of $hv/c^2$. Therefore they are not mass less for sure. *This mass is actually the energy equivalent suggested by relativity, and does not signifies the inertial mass of photons.* Second, zero mass referred by relativity should be only the inertial mass of the system, i.e.

**Photons are matter-less but not mass less.**

The case of photon provides us the distinction between matter and mass. *'Mass' can be defined as the quantity equivalent to energy of any form,* i.e. every form of energy is equivalent to some mass. The inertial energy of a body is also equivalent to some 'mass'. This inertial mass equivalent is actually *matter. Hence every lump of matter can be considered as 'mass'.* However the converse is not always true, i.e. though every matter can be regarded as mass but every mass cannot be considered as matter. Matter is actually equivalent only to the inertial energy and mass, in general, can represent equivalency to energy of any form.

A moving body possesses an inertial energy $mc^2$along with the Kinetic Energy *K*[45]. As stated in previous sections the Kinetic Energy of a moving body is actually the energy of matter wave's corpuscle associated with the body. Consequently, *the corpuscle of matter wave is actually the lump of kinetic energy of the moving-mass system, whereas the body is the lump of matter (inertial energy). The body can be said to be composed of 'matter-mass', whereas the corpuscle of matter wave as 'kinetic-mass'.* These two entities are thus physically distinct. By virtue of this distinction, the body in itself cannot be the corpuscle of the associated matter wave.

*Therefore, corpuscle is a separate entity, distinct from the body itself, in the moving matter system.*

## 13. RELATIVITY AND MATTER WAVES

In this section we have answered some theoretically posed question of relativity on the basis of the concepts of matter waves.

### 13.1. Mass of the corpuscle of matter waves

Einstein's mass energy equivalence [5,28,29,45-48] gives:

$$E' = m'c^2$$

where, $m'$ is the mass associated with the energy *E'*. As the corpuscle of matter wave carries the kinetic energy of the system, mass of the corpuscle ($m_c$) becomes:-

$$m_c = \frac{E}{c^2} = \frac{K}{c^2} = \frac{h\nu}{c^2} = \frac{(\gamma - 1)mc^2}{c^2} = (\gamma - 1)m$$

The mass of matter wave's corpuscle is a function of frequency ($\nu$) of the associated matter wave along with the velocity of body, in terms of $\gamma$.

### 13.2. Mass increment of a moving body

Answers to some concepts of relativity are associated with the concepts of matter waves, e.g. the mass increment of body with velocity [28,45,51].

Actually, it is not the mass of body which increases; rather it is the increment in the mass of the corpuscle of matter wave associated with the body, which results in the increment of the *total mass* of the system. With increasing velocity the Kinetic energy of moving mass system increases, therefore the Kinetic-mass i.e. the mass of corpuscle increases with velocity. This mass increment of the corpuscle, in reality, is responsible for the total mass increment of the system. Therefore, the total mass increment of the body is a consequence of the corpuscle's mass increment instead of the body's mass (matter).

### 13.3. Charge Invariance

When a body is in motion, its mass increases and its charge remains invariant [28]. Due to the velocity increment, the mass (kinetic-mass) of the matter wave's corpuscle increases, but the inertial mass (matter-mass) of the system still remains invariant, i.e. *the total mass of system increases (due to increase in kinetic-mass) but not the total inertial mass of system.* Since the charge of any body is associated only with its inertial mass, which remains invariant, the charge of the body also remains unaltered with velocity.

## 14. MATHEMATICAL PROOF FOR THE PROPOSED MODIFIED FREQUENCY EXPRESSION

The proposed modified frequency expression for matter waves clarifies every mathematical and conceptual inadequacies confronted by de Broglie's proposed frequency expression. In this section, a significant mathematical proof for the proposed modified frequency expression is given. We have shown that **two correct expressions (wavelength and group velocity) proposed by de Broglie contradict his own proposed frequency expression for matter waves. However, the frequency expression thus obtained satisfies the modified concept proposed in section 7, that the corpuscle of matter waves carries the kinetic energy of the moving mass system. This becomes a concrete proof for the modified frequency expression proposed in the paper.**

The proof employs the experimentally verified expression $\lambda = h/\gamma m\upsilon$ [30-36] and the conceptually correct expression for group velocity of matter waves ($\upsilon_g = \upsilon$) proposed by de Broglie [11-17]. The group velocity of a wave is [53]:-

$$\upsilon_g = \frac{d\nu}{d\left(\frac{1}{\lambda}\right)}$$

$$\Rightarrow d\nu = \upsilon_g\left[d\left(\frac{1}{\lambda}\right)\right] \qquad (23)$$

Substituting $\lambda = h/\gamma m\upsilon$ and $\upsilon_g = \upsilon$ in equation (23):-

$$d\nu = \frac{m}{h}\left\{\upsilon\left[1-\left(\frac{\upsilon}{c}\right)^2\right]^{-3/2}\right\}d\upsilon$$

Integrating back the above relation by taking proper limits, we check whether the obtained frequency expression is identical as proposed by de Broglie or not.

$$\int_0^{\nu} d\nu = \frac{m}{h}\int_0^{\upsilon}\left\{\upsilon\left[1-\left(\frac{\upsilon}{c}\right)^2\right]^{-3/2}\right\}d\upsilon \qquad (24)$$

From section 3, the wavelength of matter wave associated with the body at rest is infinite,

consequently the frequency is zero. Integrating equation (24) provides:-

$$\nu = \frac{mc^2}{h}\left[\frac{1}{\sqrt{1-\left(\upsilon/c\right)^2}} - 1\right]$$

$$\Rightarrow h\nu = \frac{mc^2}{\sqrt{1-\left(\upsilon/c\right)^2}} - mc^2$$

Or, $$h\nu = \gamma mc^2 - mc^2 = K$$

The expression for the frequency of matter waves, obtained after integration is different from de Broglie's proposed expression. Therefore, the conceptually and experimentally correct expressions, proposed by de Broglie, contradict his own proposed frequency expression for matter waves. However, *the expression for frequency thus obtained, is the proposed modified frequency expression. This satisfies the concept, that the energy of matter wave's corpuscle (hv) is equal to the kinetic energy (K) of the system, rather being equal to the total energy of the moving mass system.*

## 15. THE WAVE-PARTICLE DUALITY PICTURE

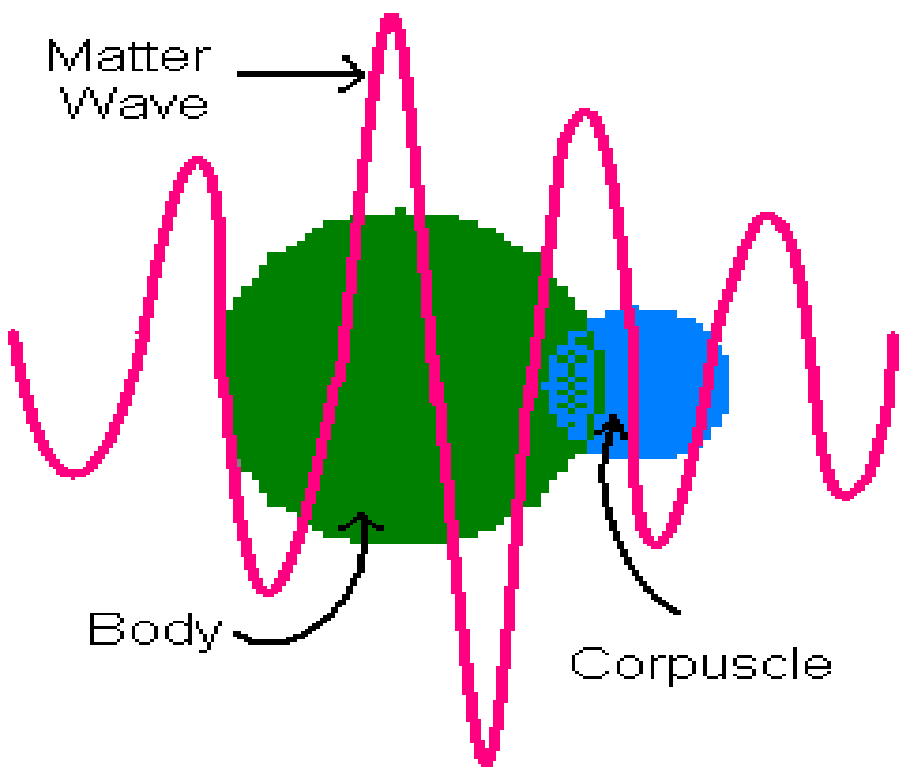

**Figure 3. Matter wave and its corpuscle associated with the body.**

Figure 3 represents the matter waves and its corpuscle associated with the body. It is presented only to offer a rough idea concerning the association of matter waves and the presence of the corpuscle of matter waves, with the body. The bigger circle in figure 3 represents a moving body. The associated matter waves are presented in form of wavy line and the corpuscle of matter wave is the smaller circle. *The corpuscle of matter wave is associated with the body, but the body in itself is not the corpuscle of the matter waves. Thus the corpuscle is an entity distinct from the body itself.*

Although the dual characteristics of radiation and matter have been verified experimentally [1-4,6,21-25,28,60,61], various questions associated with the wave-particle duality exist in the literature. In this section, elucidation to such questions is presented, based on the concepts proposed in the section 6 and 7.

**First**, Itzhak Orion and Michael Laitman [62], in their research paper regarded the particle-wave duality, as problematic. The unification of wave type character and particle type character seems very odd at the first look.

**Second**, even de Broglie, during the Nobel lecture [16], said that for radiation and matter it is necessary to introduce both the corpuscular concept and the wave concept at the same time. Again, the explanation for the unification of two seemingly incompatible and odd concepts is not known.

**Third**, in section-2.1.1, we have posed an ambiguous question: 'does body (matter) actually possess particle type character in itself? Light possesses particle nature even in the absence of matter!!!!' Possession of particle type character, by light, even in the absence of matter also needs an explanation. Moreover, the identity of a true particle should also be clarified.

**Fourth**, the wavelength and frequency expression for matter waves are:-

$$\lambda = \frac{h}{P} \text{ And } h\nu = K$$

The above two expressions relates the particle characteristic and wave characteristic of matter and radiation. To specify a wavy quantity (frequency or wavelength), the knowledge of the particle type quantity (momentum or kinetic energy) is necessary.

Both the particle type aspect and wave type aspects are mathematically interrelated.

Now the following questions arise:-

(1) Light is well known as a wave, but it still possesses particle type character. Similarly, matter is known to possess particle type character; they possess wave nature simultaneously. *But how and why do they exhibit dual character?*
(2) Why the two seemingly incompatible aspects of wave-particle duality mathematically related to each other?
(3) Why the two odd concepts of wave and particle unifies together, for matter and radiations?

*The explanation to all the questions posed above is:-*

The two SEEMINGLY different theories of waves and particles are intrinsically related to each other. These two characteristics (wave type and particle type) are inseparable. Both of them are the cause of each other. **Particle type behavior is not a characteristic of matter; rather it is a characteristic of the waves only.** Light waves possess particle nature because of its corpuscles (photons). Similarly, the particle nature exhibited by a body is because of the corpuscle of the matter waves, associated with the body; and not because of the body (matter) itself. Hence not only radiation or matter, but **each and every type of wave should exhibits a dual character.**

The justification of above concepts is:-

In previous section, the proposed modified frequency expression is proved mathematically, which supports the proposed concept that the corpuscle of matter waves, associated with the body, carries kinetic energy and the body in itself is not the corpuscle of the associated matter wave. Since, the matter waves are linked with the body, the kinetic energy, carried by the corpuscle of the matter wave, is also associated with the body, but *it is only the corpuscle of the matter waves which carries the kinetic energy* of the moving mass system. Moreover, the kinetic energy of a system represents the particle type characteristics. Therefore, particle type characteristic of matter is possessed only by the corpuscle and not by the body itself. Hence, the particle type nature observed in matters is due to the presence of the corpuscle of matter waves and not the body itself. *Therefore, particle type nature of a moving body is a result of the associated matter wave and its corpuscle.* This shows the unification of two contradictory seeming concepts (waves and particles).

Particle type nature can be observed for both matter and radiations. An Electro-magnetic radiation has particle type character due to the presence of photons. Similarly, a moving body behaves like a wave because of the associated matter waves and it interacts like a particle which is also due to the presence of matter wave's corpuscle. *This implies that both the wavy nature and the particle nature of a body is a result of the associated matter waves.* ***Body (matter) in itself has nothing to do with either of the aspect, i.e. particle or wave. It is only the associated matter waves and the corpuscle of matter waves, which gives rise to particle and wave nature of the moving body. Both radiation and matter possesses dual character because of the presence of wave. This again shows the unification of two contradictory seeming concepts (wave and Particle). The presence of any one characteristic (wave or particle), confirms the presence of the other one. Therefore, equation 13 and 15 relates the wave type variables with the particle type variables. These two seemingly odd concepts are not at all separated from each other.***

De Broglie, on the basis of symmetry of nature, proposed that if a light wave can possess particle type characteristic, then a body should also possess wave characteristic [27]. However, on the basis of the proposed interrelation of the particle and wave aspects, it becomes clear that **the body has to possess wave characteristic due to the possession of particle characteristics.**

Since, the body (matter) in-itself is independent of particle aspect, then is it possible that particle nature could be visualized even in the absence of matter? Yes, it can be visualized in case of electro-magnetic (Em) waves. *Em waves are matter less, but they still possess particle nature because the presence of particle nature depends only on the presence of wavy nature. Since Em waves are 'waves', they should necessarily possess particle type character.*

As stated above, the presence of particle type characteristic depends only on the presence of wave characteristic. *Therefore, not only radiation or matter, but each and every type of wave may exhibit a dual character. The momentum and kinetic energy of every wave should be related by the expression:-*

$$P = \frac{h}{\lambda} \textit{ And } K = h\nu$$

i.e. *the above two relations should be universally true for all waves.*

## 16. MODIFICATIONS IN THE WAVE FUNCTION EXPRESSION FOR A FREELY MOVING PARTICLE AND ITS ENERGY OPERATORS

In this section, we have shown that the correction in the frequency expression modifies the wave function and the energy operators of the freely moving particle.

### 16.1. Wave Function of a Freely Moving Particle

The existing wave function (Ψ) of a freely moving particle is [63, 64]:

$$\Psi = A\exp[\frac{-i}{\hbar}(Et - Px)] \qquad (25)$$

The wave function is usually derived from the general solution of a sinusoidal plane wave, given as [63, 65]:-

$$\Psi = \mathrm{A}\exp[-i(\omega t - \kappa x)] \qquad (26)$$

The angular frequency (*ω*) and the angular wave no. (*κ*) is then substituted in energy and momentum terms. However, as we have modified the frequency expression of matter waves, the wave function also get modifies.
From eq. [(16) and (19)]:-

$$\omega = 2\pi\nu = \frac{K}{\hbar} \text{ And } \kappa = \frac{2\pi}{\lambda} = \frac{P}{\hbar}$$

where, $\hbar = h/2\pi$ .Substituting above expressions in equation (26), the modified wave function of a freely moving particle becomes:-

$$\Psi = \mathrm{A}\exp\left(-\frac{i}{\hbar}(Kt - Px)\right) \qquad (27)$$

This is the proposed modified wave function expression for the freely moving particle. It differs from the existing wave function in energy term. Since, *hν* is equal to the kinetic energy of system instead of the total energy; therefore, the correct wave function of a freely moving particle should have only the kinetic energy term instead of the total energy term.

### 16.2. Modifications in Operators

The variation in the wave-function of a freely moving body, also introduces some changes in the dynamic quantity operators.

#### 16.2.1. Momentum operator ($\hat{P}$)

Differentiating the modified wave function equation (27) partially w.r.t. *x*:-

$$P\Psi = (-i\hbar)\frac{\partial \Psi}{\partial x}$$

The momentum operator becomes,

$$\hat{P} = (-i\hbar)\frac{\partial}{\partial x} \qquad (28)$$

The momentum operator is unchanged and is valid for both relativistic and non-relativistic velocity of the body.

#### 16.2.2. Kinetic Energy Operator ($\hat{K}$)

Differentiating (27) partially w.r.t. *'t'*, provides:-

$$K\Psi = i\hbar\frac{\partial \Psi}{\partial t}$$

Hence the kinetic energy operator becomes:-

$$\hat{K} = i\hbar\frac{\partial}{\partial t} \qquad (29)$$

This is the modified kinetic energy operator. From the earlier wave function of freely moving body (eq. 25), $i\hbar\frac{\partial}{\partial t}$ is regarded as the total energy operator [28,29,59,66-69]. However, the modified wave function reveals it as a kinetic energy operator. This kinetic energy is valid for both relativistic and non-relativistic cases.

*If the particle is not acted upon by any external force i.e. potential energy acting on it is zero, then the total energy becomes equal to the kinetic energy. Therefore the kinetic energy operator becomes a total energy operator for a freely moving body.*

As for the non-relativistic cases:-

$$K_n = p^2/2m \tag{30}$$

where, $K_n$ is the non-relativistic kinetic energy operator. Substituting the momentum operator (eq. 28), the non-relativistic kinetic energy operator $K_n$ becomes:-

$$\hat{K}_n = \frac{\left(-i\hbar\frac{\partial}{\partial x}\right)^2}{2m} = \frac{-\hbar^2}{2m}\frac{\partial^2}{\partial x^2} \tag{31}$$

The non relativistic kinetic energy operator is found to be unchanged, because the momentum operator does not vary [28].

Both equations (31) and (29) are the Kinetic energy operators. Therefore, they are equivalent for the non-relativistic velocity of body, i.e. for non-relativistic cases:-

$$i\hbar\frac{\partial}{\partial t} = \left[\frac{-\hbar^2}{2m}\frac{\partial^2}{\partial x^2}\right]$$

Operating the wave function $\Psi$ both sides, we get:-

$$i\hbar\frac{\partial \Psi}{\partial t} = \left[\frac{-\hbar^2}{2m}\frac{\partial^2 \Psi}{\partial x^2}\right] \tag{32}$$

This provides the Schrödinger's equation for a body, moving freely with non-relativistic velocity [28,29,58,64-71]. It is derived merely by equivalence of the kinetic energy operators, for the non-relativistic cases.

#### 16.2.3. Total energy operator ( $\hat{H}$ )

Total energy operator is defined as the Hamiltonian operator ( $\hat{H}$ ), defined as:-

Total energy = Kinetic energy + Potential energy

$$\Rightarrow \hat{H} = \hat{K} + \hat{U}$$

Substituting equation (29):-

$$\hat{H} = i\hbar\frac{\partial}{\partial t} + U(x) \tag{33}$$

Equation (33) is the proposed total energy operator, valid for both the relativistic and non-relativistic cases.
However, for non-relativistic case, kinetic energy operator becomes ( $\frac{-\hbar^2}{2m}\frac{\partial^2}{\partial x^2}$ )

$$\Rightarrow \hat{H}_n = (-\frac{\hbar^2}{2m})\frac{\partial^2}{\partial x^2} + U(x) \tag{34}$$

Therefore, the non-relativistic total energy operator ($H_n$), remains unchanged.

## 17. APPLICATIONS OF THE MODIFIED WAVE FUNCTION AND OPERATORS

In this section, we have shown that the corrected wave function and the energy operators can be implied to derive the fundamentally correct Schrödinger's equations, Klein-Gordon's Equation and Dirac's equation. Their satisfactory derivation offers a verification step for the proposed modifies wave function and energy operators of a freely moving particle.

### 17.1. Schrödinger's equation for a freely moving particle

Differentiating the corrected wave function {equation (27)} w.r.t.'*t'* partially, provides:-

$$K\Psi = i\hbar\frac{\partial \Psi}{\partial t} \tag{35}$$

Simultaneously, differentiating $\Psi$ twice partially w.r.t. *x*, offers:-

$$P^2\Psi = \left(-\hbar^2\right)\frac{\partial^2\Psi}{\partial x^2} \tag{36}$$

As for the non-relativistic cases:-

$$K = \frac{P^2}{2m} \Rightarrow K\Psi = \frac{P^2\Psi}{2m}$$

Substituting the equation [(35) and (36)] above equation gives:-

$$i\hbar\frac{\partial\Psi}{\partial t} = \frac{\left(-\hbar^2\right)}{2m}\frac{\partial^2\Psi}{\partial x^2} \tag{37}$$

Equation (37) is the Schrödinger's equation for a body moving freely at non-relativistic velocity [28,29,51,58]. Similarly, the Time-Independent form of Schrödinger's equation from the corrected wave function is found to be:-

$$\frac{\partial^2\psi}{\partial x^2} + \frac{2m}{\hbar^2}\left(K\psi\right) = 0 \tag{38}$$

Here, $\psi$ is the position dependent wave function of the body. If the particle is acted upon by a potential energy function *U*, total energy of the particle becomes:-

$$E=K+U$$

$$\Rightarrow \quad K=E-U$$

Substituting the above relation in eq. (38), the time-independent form of Schrödinger's equation becomes:-

$$\frac{d^2\psi}{dx^2} + \frac{2m}{\hbar^2}\left(E-U\right)\psi = 0 \tag{39}$$

Thus the time independent Schrödinger's equation is derived satisfactorily. It is noticed here, that the modifications in the matter wave's frequency expression and the wave function of a freely moving particle, does not modifies the existing non relativistic quantum mechanics because the Schrödinger's equations still remains invariant.

### 17.2. Klein-Gordon's and Dirac's equation

The Klein Gordon's equation relates the total energy and the relativistic momentum of a freely moving body. It is derived in this section, by using the modified wave function:

$$\Psi = \mathrm{A}\exp\left(-\frac{i}{\hbar}(Kt - Px)\right)$$

Theory of relativity provides relation among the momentum and the total energy of a body, [28, 29,45-50] as:-:

$$E^2 = \left(mc^2\right) + \left(Pc\right)^2 \tag{40}$$

In section 16.2.2, we proposed that for a freely moving body (in the absence of potential energy), the kinetic energy operator becomes total energy operator; i.e. the energy and momentum operator for a freely moving body is:-

$$\hat{E} = i\hbar\frac{\partial}{\partial t} \text{ and } \hat{P} = \left(-i\hbar\right)\frac{\partial}{\partial x}$$

Substituting both the operators in equation (40) we get:-

$$[i\hbar\frac{\partial}{\partial t}]^2 = \left(mc^2\right)^2 + \left(-i\hbar c\right)^2\frac{\partial^2}{\partial x^2}$$

$$\Rightarrow -\hbar^2\frac{\partial^2\Psi}{\partial t^2} = -\hbar^2c^2\frac{\partial^2\Psi}{\partial x^2} + m^2c^4\Psi = 0 \tag{41}$$

Equation (41) is the well known Klein Gordon's equation [74-77]. It is derived using the corrected wave function for a freely moving particle and its energy operators.

*Similarly,* the equivalency of Kinetic energy and Total energy operator (for a free particle) provides the satisfactory derivation of Dirac's Equation [78-81], given as:-

$$i\hbar\frac{\partial\Psi}{\partial t} = c\alpha\cdot P\Psi + \beta' mc^2\Psi = 0$$

where, $\alpha$ and $\beta$ all are the Dirac Matrices [66].

## 18. MATHEMATICAL PROOF FOR THE MODIFIED PHASE VELOCITY EXPRESSION

In this section, it is shown that Schrödinger's equation is just a modified form of the universal wave equation [53-55] given as:-

$$\frac{\partial^2\Psi}{\partial x^2}=\frac{1}{\upsilon_P^2}\frac{\partial^2\Psi}{\partial t^2}$$

Since this form of wave equation involves the phase velocity term, therefore, the substitution of correct $\upsilon_P$ expression should yield the time dependent Schrödinger's equation. It is shown that the substitution of de Broglie's proposed phase velocity expression fails to yield the correct result; however, the proposed modified phase velocity expression succeeds in the task. Thus, the process becomes as a proof for the correctness of the proposed phase velocity expression for matter waves. The time-independent Schrödinger's equation is:-

$$\frac{\partial^2\Psi}{\partial x^2}+\frac{2m}{\hbar^2}(E-U)\Psi=0$$

$$\Rightarrow\frac{\partial^2\Psi}{\partial x^2}+\frac{P^2}{\hbar^2}\Psi=0$$

As, $P=\hbar k$. Above equation becomes:-

$$\frac{\partial^2\Psi}{\partial x^2}+k^2\Psi=0$$

It is known that the angular wave no. of a wave is angular frequency over the wave velocity [54,55]. Therefore, the equation becomes:-

$$\frac{\partial^2\Psi}{\partial x^2}+\frac{\omega^2}{\upsilon_P^2}\Psi=0$$

From equation (26), it can be deduced that:-

$$\omega^2\Psi=-\frac{\partial^2\Psi}{\partial t^2}$$

$$\Rightarrow\frac{\partial^2\Psi}{\partial x^2}=\frac{1}{\upsilon_P^2}\frac{\partial^2\Psi}{\partial t^2}$$

Therefore, Schrödinger's equation is actually an expanded form of the universal wave equation. Here, the transformation of Schrödinger's equation into the universal wave equation is presented without involving any proposed modifications. Thus, the Schrödinger's equation transformation into general wave equation is an authentic result.

This result is now used to provide proof for the proposed phase velocity expression. The substitution of the proposed non-relativistic phase velocity expression {equation (22)} in the general wave equation provides:-

$$\frac{\partial^2\Psi}{\partial t^2}=\frac{(m\upsilon)^2}{4m^2}\frac{\partial^2\Psi}{\partial x^2}$$

As, for non-relativistic cases [66,73] $(m\upsilon)^2=2mK$

$$\Rightarrow\frac{\partial^2\Psi}{\partial t^2}=\frac{2m}{4m^2}(K)\frac{\partial^2\Psi}{\partial x^2}$$

Substituting the kinetic energy operator {equation (29)} provides:-

$$\frac{\partial}{\partial t}\left(\frac{\partial\Psi}{\partial t}\right)=\left(\frac{i\hbar}{2m}\right)\frac{\partial}{\partial t}\left(\frac{\partial^2\Psi}{\partial x^2}\right)$$

$$\Rightarrow i\hbar\frac{\partial\Psi}{\partial t}=(-)\frac{\hbar^2}{2m}\frac{\partial^2\Psi}{\partial x^2}$$

The obtained equation is the time dependent Schrödinger's equation and is achieved by the substitution of modified phase velocity expression in the universal wave equation. **However, the same result is not obtained if the process is followed by using de Broglie's proposed phase velocity expression ($c^2/\upsilon$).** This shows the appropriateness of the proposed modified phase velocity expression over de Broglie's proposed phase velocity expression for matter waves.

## 19. NOVEL RELATION BETWEEN KINETIC ENERGY AND MOMENTUM OF A BODY

In this section, we have presented a general relation between kinetic energy and momentum of a moving mass system. From equation (16) and (19), we have:-

$$\frac{1}{\lambda}=\frac{P}{h};\nu=\frac{K}{h}$$

$$\Rightarrow K = P(\lambda \nu) = P\upsilon_p \qquad (42)$$

where, $\upsilon_p$ is the modified phase velocity of matter waves associated with the moving body, given by equation (20):-

$$\upsilon_P = \frac{c^2}{\upsilon}\left[1 - \sqrt{1 - \frac{\upsilon^2}{c^2}}\right]$$

Equation (42) is the proposed *general relation between kinetic energy and momentum of the moving body, valid for both relativistic and non-relativistic cases.*

Equation (42) can also be reduced to the non-relativistic Kinetic Energy-Momentum relation ($K=P^2/2m$). Since, for the non-relativistic velocity ($\upsilon$) of the body, the phase velocity ($\upsilon_p$) of matter waves becomes $\upsilon/2$ [from eq. (22)], and the momentum becomes $m\upsilon$. Equation (42) can therefore be written as:-

$$K = \frac{(m\upsilon)\upsilon}{2} = \frac{(m\upsilon)(m\upsilon)}{2m} = \frac{P_n^2}{2m}$$

where, $P_n$ is the non relativistic momentum of the body.

Therefore, the proposed Kinetic energy-Momentum relation is consistent with the non-relativistic relation. The relation, in upcoming sections, is utilized for the analysis of Quantum Mechanical Systems, in relativistic domain. The use of relativistic kinetic energy and momentum, obtained from the special theory of relativity [45-50]:-

$$K = \sqrt{(Pc)^2 + (mc^2)^2} - mc^2$$

introduces mathematical complications in analysis of relativistic Quantum Mechanical Systems. Therefore use of new relation [eq. (42)], help us to analyze the situations in a convenient manner. The proposed relation is also employed to derive the general and relativistic Quantum Mechanical Wave equations.

## 20. INTRODUCTION OF NOVEL QUANTUM MECHANICAL WAVEEQUATIONS- GENERAL AND RELATIVISTIC

In this section, we have proposed novel General and Relativistic Quantum Mechanical Wave Equations.

### 20.1. General quantum mechanical wave equations

In this section, we have derived the general quantum mechanical wave equations, valid for both non-relativistic and relativistic cases. These equations allow us to analyze the particle's behavior even at relativistic velocity.

#### 20.1.1. Time Dependent Form

The Kinetic Energy and Momentum of the moving body are related, according to the general relation, given by equation (42):-

$$K = (P)(\upsilon_p)$$

**This relation is used to derive the General Quantum Mechanical Wave equation**. Operating the wave function ($\Psi$) both sides of the equation gives:-

$$K\Psi = (\upsilon_p)P\Psi \qquad (43)$$

Substituting the Momentum and Kinetic energy operators {eq. (28) and (29)}:-

$$i\hbar\frac{\partial \Psi}{\partial t} = (-\upsilon_P)\left(i\hbar\frac{\partial \Psi}{\partial x}\right) \qquad (44)$$

$$i\hbar\frac{\partial \Psi}{\partial t} = (-\upsilon_P)(i\hbar\nabla \bullet \Psi)$$

Equation (44) is the proposed General Quantum Mechanical Wave Equation, in time dependent form. Since it is the general wave equation, it is valid for both relativistic and non-relativistic cases. It is first order, in both space and time coordinates; similar to Dirac's equation [78-81] for spin-1/2 particle, and thus satisfies the space-time symmetry requirement of relativity [66]. This equation is dependent on the phase velocity term, therefore on the velocity of body. This general wave equation can be reduced to both Schrödinger's and Dirac's equation, by the substitution of phase velocity expression in terms of kinetic energy and momentum of body according to the situation.

Since it is the general wave equation, it can be transformed into the Schrödinger's equation as follows:-

For the non-relativistic cases, the phase velocity term $\upsilon_p$ becomes half the velocity of particle, from equation (22). Therefore, the general Wave Equation (44), for the non-relativistic cases becomes:-

$$i\hbar\frac{\partial\Psi}{\partial t}=\left(-\frac{\upsilon}{2}\right)\left(i\hbar\frac{\partial\Psi}{\partial x}\right)$$

$$\Rightarrow i\hbar\frac{\partial\Psi}{\partial t}=\frac{(m\upsilon)}{(-2m)}\left[i\hbar\frac{\partial\Psi}{\partial x}\right]$$

$$\Rightarrow i\hbar\frac{\partial\Psi}{\partial t}=\frac{P}{-2m}\left[i\hbar\frac{\partial\Psi}{\partial x}\right]$$

Substitution of the momentum operator {equation (28)} in above equation provides:-

$$i\hbar\frac{\partial\Psi}{\partial t}=\left(\frac{-\hbar^2}{2m}\right)\frac{\partial^2\Psi}{\partial x^2}$$

The equation obtained above is the time dependent Schrödinger's equation [sec. XVII.A]. The satisfactory derivation of Schrödinger's equation from the proposed General Wave Equation (44), demonstrates the accuracy of the proposed equation.

The general wave equation (44) can also be transformed into Dirac's equation, for the relativistic cases, as follows:-

Since, $\upsilon_P=K/P$ form equation (44). Therefore, the general wave equation becomes:-

$$i\hbar\frac{\partial\Psi}{\partial t}=\left(-\frac{K}{P}\right)\left(i\hbar\frac{\partial\Psi}{\partial x}\right)$$

From Special Theory of Relativity, we have [28]:-

$$P=\frac{\sqrt{K^2+2mKc^2}}{c}$$

$$\Rightarrow i\hbar\frac{\partial\Psi}{\partial t}=\frac{Kc}{\sqrt{K^2+2mKc^2}}\left(i\hbar\frac{\partial\Psi}{\partial x}\right)$$

$$\Rightarrow \frac{\partial^2\Psi}{\partial t^2}=\frac{K^2c^2}{K^2+2mKc^2}\frac{\partial^2\Psi}{\partial x^2}$$

$$\Rightarrow \frac{\partial^2\Psi}{\partial t^2}=\frac{Kc^2}{K+2mc^2}\frac{\partial^2\Psi}{\partial x^2} \qquad (45)$$

$$\Rightarrow (K+2mc^2)\frac{\partial^2\Psi}{\partial t^2}=Kc^2\frac{\partial^2\Psi}{\partial x^2}$$

Substituting the Kinetic energy operator {equation (29)} in above equation provides:-

$$\frac{\partial}{\partial t}\left(i\hbar\frac{\partial^2\Psi}{\partial t^2}+2mc^2\frac{\partial\Psi}{\partial t}\right)=i\hbar c^2\frac{\partial}{\partial t}\left(\frac{\partial^2\Psi}{\partial x^2}\right)$$

$$\Rightarrow i\hbar\frac{\partial^2\Psi}{\partial t^2}+2mc^2\frac{\partial\Psi}{\partial t}=i\hbar c^2\left(\frac{\partial^2\Psi}{\partial x^2}\right)$$

$$\Rightarrow (i\hbar)^2\frac{\partial^2\Psi}{\partial t^2}+2mc^2(i\hbar)\frac{\partial\Psi}{\partial t}=(i\hbar)^2c^2\frac{\partial^2\Psi}{\partial x^2}$$

$$\Rightarrow \left[i\hbar\frac{\partial}{\partial t}+mc^2\right]^2\Psi=\left[\left(-i\hbar\frac{\partial}{\partial x}\right)^2c^2+m^2c^4\right]\Psi$$

Since $\left(-i\hbar\frac{\partial}{\partial x}\right)$ is the momentum operator {equation (28)}

$$\Rightarrow \left[i\hbar\frac{\partial}{\partial t}+mc^2\right]^2\Psi=\left[P^2c^2+m^2c^4\right]\Psi$$

$$\Rightarrow \left(i\hbar\frac{\partial}{\partial t}+mc^2\right)\Psi=c\left[\alpha\cdot P+\xi mc\right]\Psi$$

$$\Rightarrow \frac{i\hbar\partial\Psi}{\partial t}=c\alpha\cdot P\Psi+(\xi-1)mc^2\Psi$$

Above equation is similar to Dirac's equation, by assuming $(\xi\text{-}1)=\beta'$. Therefore, the equation becomes Dirac's equation, given as:-

$$i\hbar\frac{\partial\Psi}{\partial t}=c\alpha\cdot P\Psi+\beta' mc^2\Psi$$

Thus, Dirac's equation can be derived from the general wave equation (44).

*20.1.1.1. Interpretation of the wave function obtained from the General Wave equation.* The quantity $\left|\Psi(r,t)\right|^2$, obtained from Schrödinger's and Dirac's equation, is defined as the position probability density *P(r.t)* [64,66,74,82,83] In this section, it is illustrated that the proposed general wave equation also agrees with this definition of position probability density. The general wave equation satisfies the common form of continuity equation and therefore, the square of the wave function symbolizes the position probability density of the particle. The proposed general wave equation is:-

$$i\hbar\frac{\partial\Psi}{\partial t} = (-i\hbar)\upsilon_p(\nabla\bullet\Psi)$$

Multiplying above equation by Ψ* and the complex conjugate of above equation by Ψ; then subtracting second from first provides:-

$$i\hbar\left[(\Psi^*\frac{\partial\Psi}{\partial t})+(\Psi\frac{\partial\Psi^*}{\partial t})\right]=(-i\hbar\upsilon_p)[(\Psi^*\nabla\bullet\Psi)+(\Psi\nabla\bullet\Psi^*)]$$

$$\frac{\partial(\Psi^*\Psi)}{\partial t} = (-\upsilon_p)[\nabla\bullet(\Psi^*\Psi)]$$

$$\Rightarrow \frac{\partial(\Psi^*\Psi)}{\partial t} + \nabla\bullet[\upsilon_p(\Psi^*\Psi)] = 0$$

Comparison of the above equation with the general continuity equation [66,84]:-

$$\frac{\partial[\rho(r.t)]}{\partial t} + \nabla\bullet[j(r,t)] = 0$$

where, *ρ(r.t)* can be interpreted as the probability density and *j(r.t)* is the probability current density, offers:-

$$\rho(r,t) = \Psi^*\Psi \text{ and } j(r,t) = \upsilon_p(\Psi^*\Psi)$$

Therefore, the quantity Ψ*Ψ represents the position probability density of the particle and $[\upsilon_p(\Psi^*\Psi)]$ represents the probability current density.

### 20.1.2 Time Independent Form

Many situations are possible, when the particle in a Quantum Mechanical system attains relativistic velocity; For e.g. a diatomic molecule, equivalent to a rigid rotor, rotating at relativistic velocity, a particle trapped in a potential well of sufficiently small length attains a relativistic velocity, a hydrogen type atom, with sufficiently high no. of protons, so that the motion of electron in the vicinity of nucleus may corresponds to the relativistic velocity domain etc. Though relativity considers space-time to be inseparable, the consideration of only space coordinate is sufficient even for such relativistic situations, due to the dependence of potential acting on particle explicitly on the space coordinate. Consequently, a General Quantum Mechanical Wave Equation only in space coordinate is required. It is derived in this section. The application of the equation in quantum mechanical systems provides the information regarding the particle's behavior at any velocity (relativistic or non-relativistic). Equation (27), the wave function of a freely moving particle is given as:-

$$\Psi = \mathrm{A}\exp\left[\frac{-i}{\hbar}(Kt - Px)\right]$$

Differentiating it twice, partially w.r.t. '*x*':-

$$\frac{\partial^2\Psi}{\partial x^2} + \frac{P^2}{\hbar^2}\Psi = 0 \qquad (46)$$

Let the wave function Ψ be represented as product of position dependent and time dependent wave function. i.e.:-

$$\Psi = \psi(x)\times f(t)$$

So, eq. (46) becomes:-

$$\frac{\partial^2\psi}{\partial x^2} + \frac{P^2}{\hbar^2}\psi = 0$$

Substituting $P=K/\upsilon_P$, from equation (42) in above equation gives:-

$$\frac{\partial^2\psi}{\partial x^2} + \frac{K^2}{\hbar^2\upsilon_p^2}\psi = 0 \qquad (47)$$

If the potential energy function acting on the particle is *U*, then the total energy (*E*) of the particle is the sum of its kinetic energy and potential energy:-

$$E = K + U$$
$$\Rightarrow K = E - U$$

Substituting the above kinetic energy expression in equation (47) provides:-

$$\frac{d^2\psi}{dx^2}+\frac{(E-U)^2}{\hbar^2\upsilon_p^2}\psi=0 \qquad (48)$$

Equation (48) is the proposed General Wave Equation in time-independent form. It can also be transformed to the time independent Schrödinger's equation as follows:

Since, for non-relativistic cases:-

$$\upsilon_P^2=\left(\frac{\upsilon}{2}\right)^2=\left(\frac{P_n}{2m}\right)^2=\frac{K_n}{2m}$$

where, $P_n$ and $K_n$ are the non-relativistic kinetic energy and momentum terms. Hence, equation (48) becomes:-

$$\frac{d^2\psi}{dx^2}+\frac{P_n^{\ 2}}{\hbar^2}\psi=0 \qquad (49)$$

As for non-relativistic cases:-

$$P_n^{\ 2}=2mK_n=2m(E-U)$$

Hence, equation (49) becomes:-

$$\frac{d^2\psi}{dx^2}+\frac{2m(E-U)}{\hbar^2}\psi=0$$

This is the well known Time-Independent Schrödinger's equation [63-73,85-87] and is derived from the proposed general wave equation(48).A point worth to be noted here is that the general form of the proposed time-independent wave equation is same, as that of Schrödinger's Equation, i.e.

$$\frac{d^2\psi}{dx^2}+\kappa^2\psi=0$$

is the general form of both the time independent wave equations. Here, κ is the angular wave number for matter waves. i.e.:-

$$\kappa=\frac{2\pi}{\lambda}=2\pi\frac{P}{h}$$

Moreover the eq. (48) satisfies the $\hat{H}\psi=E\psi$ form as well. Equation (48) can be written as:-

$$\frac{\partial^2\psi}{\partial x^2}=(-)\frac{(E-U)^2}{\hbar^2\upsilon_p^2}\psi$$

$$\Rightarrow (E-U)\psi=(-i\hbar\upsilon_P)\frac{\partial\psi}{\partial x}$$

Substituting equation (44) in it:-

$$(E-U)\psi=i\hbar\frac{\partial\psi}{\partial t}$$

$$\Rightarrow E\psi=i\hbar\frac{\partial\psi}{\partial t}+U\psi$$

$$\Rightarrow \hat{H}\psi=E\psi$$

Hence the General wave equation satisfies this requirement also, similar to Schrodinger's [28] and Dirac's equation [66].

Since equation (48) is general wave equation therefore it permits the determination of particle's behavior in situations even at relativistic velocity. The resulting relativistic wave equation is derived in the upcoming section.

### 20.2 Relativistic Wave Equation

In this section, we have proposed a time independent Relativistic Quantum Mechanical Wave Equation by using the general wave equation (48). The substitution of non-relativistic Kinetic energy-Momentum relation at place of phase velocity term provided Schrodinger's equation, in previous section. Similarly, the substitution of a relativistic Kinetic energy and Momentum relation at place of phase velocity term, introduces a Relativistic Wave equation. Equation (48) is:-

$$\frac{d^2\psi}{dx^2}+\frac{(E-U)^2}{\hbar^2\upsilon_p^2}\psi=0$$

As $K=P\upsilon_P$, from equation (42). Therefore, the above equation becomes:-

$$\frac{d^2\psi}{dx^2}+\frac{P^2}{\hbar^2}\psi=0$$

The substitution of non-relativistic Kinetic energy-Momentum relation in above equation provides us the Schrodinger's equation (previous section). Similarly the substitution of momentum, in terms of relativistic Kinetic energy provides a Relativistic Wave Equation. Special theory of relativity offers a relation between the relativistic Kinetic energy and Momentum of the particle as [28]:-

$$P^2=\frac{K^2+2mc^2K}{c^2}$$

The substitution of this relation provides relativistic wave equation as:

$$\frac{d^2\psi}{dx^2}+\frac{K^2+2mc^2K}{\hbar^2c^2}\psi=0$$

As, $E=K+U$. Therefore, above equation becomes:-

$$\frac{d^2\psi}{dx^2}+\frac{(E-U)^2+2mc^2(E-U)}{\hbar^2c^2}\psi=0 \quad (50)$$

$$\frac{d^2\psi}{dx^2}+\frac{2m(E-U)}{\hbar^2}+\frac{(E-U)^2}{\hbar^2c^2}\psi=0 \quad (51)$$

$$\frac{d^2\psi}{dx^2}+\frac{2m(E-U)}{\hbar^2}\left[1+\frac{(E-U)}{2mc^2}\right]\psi=0 \quad (52)$$

Above equation is the proposed Relativistic time-independent wave equation. The simple addition of the term $(E-U)^2/\hbar^2c^2$, in the Schrödinger's equation yields out the relativistic wave equation (51). From equation (52), it is observed that for $K=(E-U)<<mc^2$, the quantity inside bracket becomes almost unity, yielding Schrödinger's equation directly for the non-relativistic cases.

The proposed Relativistic Wave equation (51) is derived by using the general wave equation (48). Similarly, we have also derived Schrödinger's equations and Dirac's equation from it. Therefore, all these equations are interrelated with the General wave equations (44) and (48). This interrelation is shown in figure 4.The Schrödinger's, Dirac's and the introduced relativistic quantum mechanical wave equations are distinctly obtainable from the proposed general quantum mechanical wave equation (44) and (48).

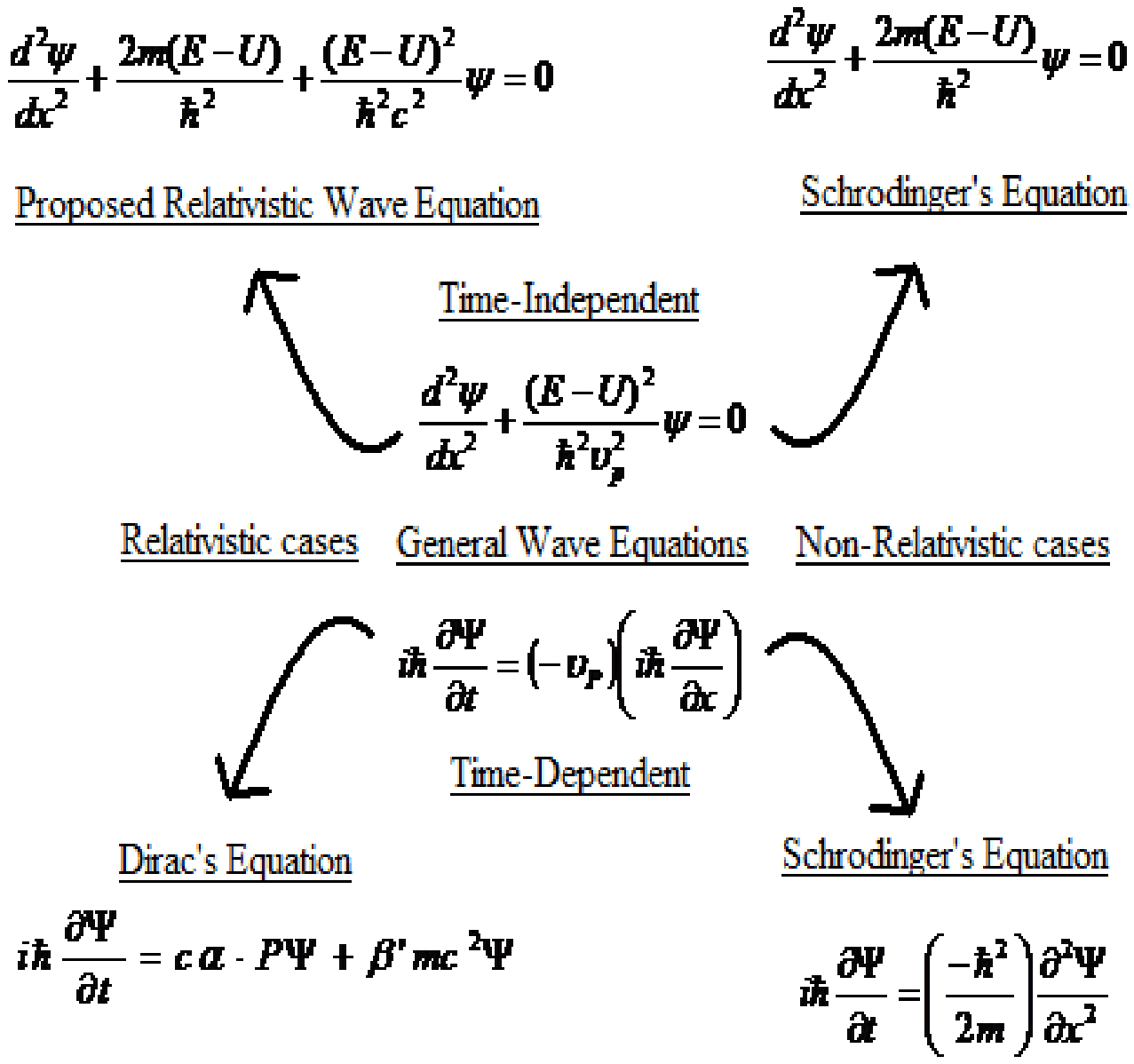

**Figure 4 Pictorial representation of the interrelation of the Wave Equations.**

Figure 4 depicts that for non-relativistic cases the Schrödinger's equations are derivable from the general wave equations. Similarly, for the relativistic cases, Dirac's equation and the proposed relativistic wave equation (51) can also be obtained from the general wave equations. Thus, all the quantum mechanical wave equations can be obtained from general wave equation (44) and (48).

## 21. SINGLE WAVE EQUATION FOR ALL

In this section, it is presented that the proposed general and relativistic wave equations, along with Dirac's equation, follows the universal wave equation; similar to Schrödinger's equation {section 18.1}. Consequently, all the six wave equations {two Schrodinger's, one Dirac's and three proposed} are distinctly obtainable from the universal wave equation itself. The universal wave equation is:-

$$\frac{\partial^2\Psi}{\partial x^2}=\frac{1}{\upsilon_P^2}\frac{\partial^2\Psi}{\partial t^2} \qquad (53)$$

### 21.1 Dirac's equation

We shall first show the conversion of universal wave equation (53) into Dirac's equation. For that we can substitute $\upsilon_P=K/P$, from equation (42), into the universal wave equation (53). Therefore:-

$$\frac{\partial^2\Psi}{\partial x^2}=\frac{P^2}{K^2}\frac{\partial^2\Psi}{\partial t^2} \qquad (54)$$

Substituting the relation for relativistic cases [28]:-

$$P^2=\frac{K^2+2mKc^2}{c^2}$$

in equation (54) provides:-

$$\frac{\partial^2\Psi}{\partial x^2}=\frac{K^2+2mKc^2}{c^2K^2}\frac{\partial^2\Psi}{\partial t^2}$$

$$\Rightarrow\frac{Kc^2}{K+2mc^2}\frac{\partial^2\Psi}{\partial x^2}=\frac{\partial^2\Psi}{\partial t^2} \qquad (55)$$

Equation (55) is identical to equation (45). The further conversion of it into Dirac's equation is already shown in section (22.1.1). Consequently, Dirac's equation is obtained from the universal wave equation itself, given as:-

$$i\hbar\frac{\partial\Psi}{\partial t}=c\alpha\cdot P\Psi+\beta' mc^2\Psi$$

Therefore, Dirac's equation is an expanded form on the universal wave equation (53).

### 21.2 Proposed general and relativistic wave equation

The proposed general {equation (44) and (48)} and relativistic wave equations is now shown to be obtainable from the universal wave equation (53). Taking square root of the universal wave equation (53):-

$$\frac{\partial\Psi}{\partial t}=\pm\upsilon_P\frac{\partial\Psi}{\partial x}$$

The two equations thus obtained are same; the positive sign indicates the wave travelling on positive x-axis and vice versa [53]. Therefore:-

$$\frac{\partial\Psi}{\partial t}=-\upsilon_P\frac{\partial\Psi}{\partial x}$$

$$\Rightarrow i\hbar\frac{\partial\Psi}{\partial t}=-i\hbar\upsilon_P\frac{\partial\Psi}{\partial x}$$

Above equation is the proposed general wave equation (44) in time dependent form. It is obtained from the universal wave equation. Similarly, to obtain the general wave equation in time independent form, the universal wave equation by the use of $\upsilon_P=\omega/k$ [28,53] becomes:-

$$\omega^2\frac{\partial^2\Psi}{\partial x^2}=k^2\frac{\partial^2\Psi}{\partial t^2}$$

From equation (26), representing a plane wave equation, it can be deduced that:-

$$\omega^2\Psi=-\frac{\partial^2\Psi}{\partial t^2}$$

$$\Rightarrow\frac{\partial^2\Psi}{\partial x^2}+k^2\Psi=0$$

Since for matter wave, $P=\hbar k$. Above equation becomes:-

$$\frac{\partial^2\Psi}{\partial x^2}+\frac{P^2}{\hbar^2}\Psi=0 \qquad (56)$$

From equation (42), $P=K/\upsilon_P=(E-U)/\upsilon_P$. Therefore, equation (56) becomes:-

$$\frac{\partial^2\Psi}{\partial x^2}+\frac{(E-U)^2}{\hbar^2\upsilon_P^2}\Psi=0$$

Above equation is the proposed general wave equation (48) in time independent form. Similarly, the substitution of relation $P^2=(K^2+2mKc^2)/c^2$, provided by relativity [28], in equation (56) provides:-

$$\frac{\partial^2\Psi}{\partial x^2}+\frac{(E-U)^2+2m(E-U)c^2}{\hbar^2c^2}\Psi=0$$

The above equation is also the proposed relativistic wave equation (51).

Consequently, all the proposed wave equations {equation (44), (48) and (51)}, along with Dirac's equation, are shown as the expanded forms of the universal wave equation, under different substitutions.

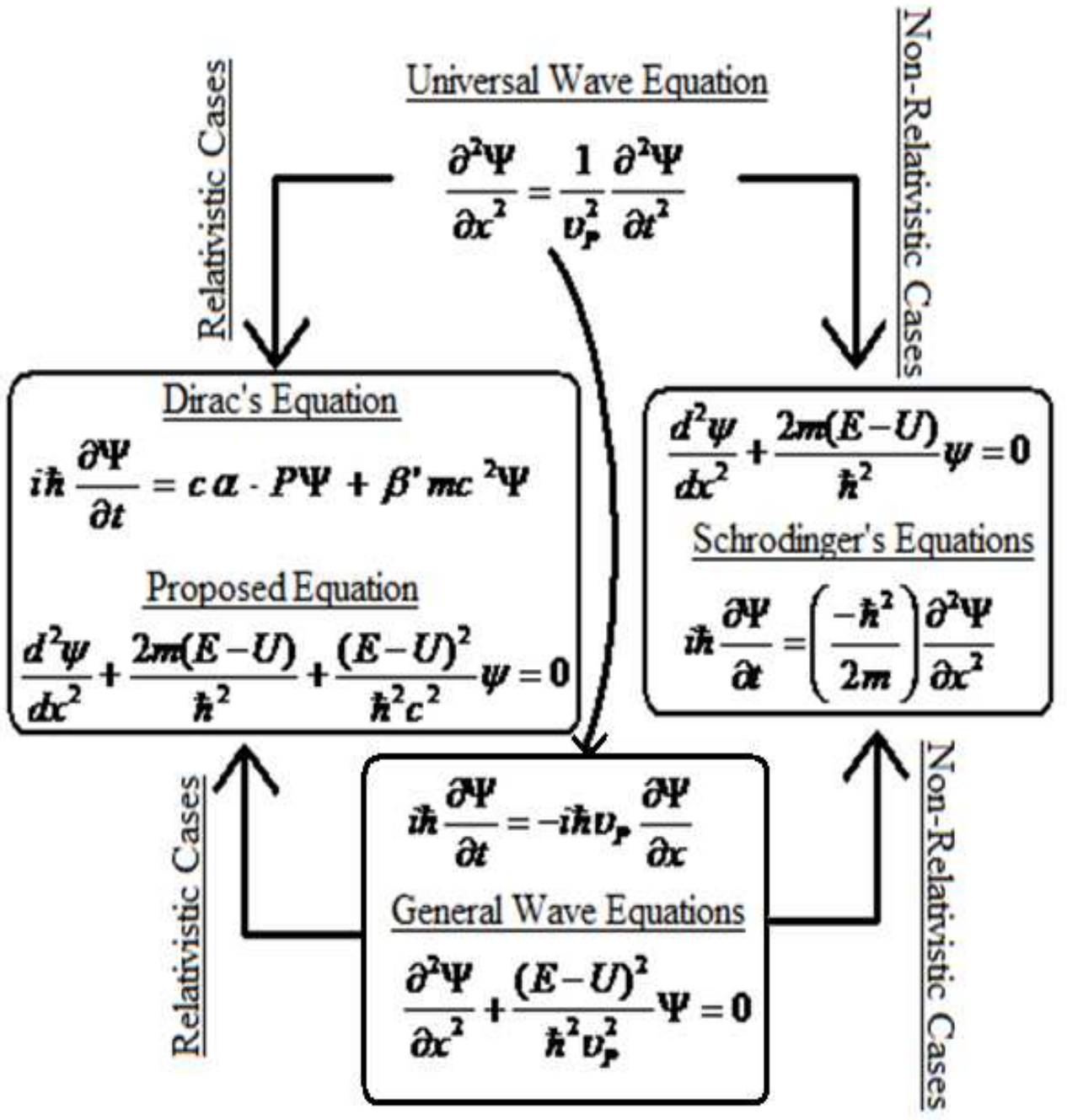

**Figure 5 Pictorial Representation of the Quantum Mechanical Wave Equation's interrelation with the Universal Wave Equation.**

Therefore, this section and section 17 shows that the all the quantum mechanical wave equations, including Schrodinger's equation, are the different expanded forms of the single Universal Wave equation. Figure 5 explains this pictorially. Figure 5 depicts the interrelation of the wave equations with the universal wave equation. **All the six wave equations are obtainable from the single universal wave equation itself.** The proposed General Quantum Mechanical Wave Equation can also be employed to derive Schrodinger's and Dirac's equations. The proposed Relativistic Quantum mechanical wave equation also emerges from the proposed general wave equations. Consequently, the derivation of proposed equations (44), (48) and (51) from the universal wave equation provides a step for the reliability on the results that would be obtained by the application of the proposed equations.

In the upcoming sections, the proposed equations are applied to some important quantum mechanical systems to extract the knowledge regarding particle's behavior in such systems, even at relativistic velocity.

## 22. PARTICLE TRAPPED IN A BOX – GENERAL AND RELATIVISTIC

In this section, a simple approach to the theoretical situation of particle trapped in an infinitely hard box is presented. The general energy Eigen values for the particle is obtained and is subsequently converted into the non-relativistic expression. By utilizing the obtained general energy Eigen values expression the energy Eigen values for the relativistic cases is determined. Consider a particle of inertial mass '*m*' moving with a velocity '*v*'; trapped in a box of infinitely hard walls, separated by a distance '*L*'. The particle goes back and forth in between the walls and thus the associated matter waves forms a standing wave pattern between the walls of the box [28]. Therefore, the length of the box '*L*' is an integral multiple of half the de Broglie wavelength ($\lambda/2$) [28] i.e.:-

$$L=\frac{n\lambda}{2}$$

As, $P=h/\lambda$ and $\lambda=2L/n$, the momentum ($P$) becomes:-

$$P=\frac{nh}{2L} \qquad (57)$$

### 22.1 For relativistic cases

Theory of relativity provides the relation [48]:-

$$E = \sqrt{(Pc)^2 + (mc^2)^2}$$
$$\Rightarrow K = \sqrt{(Pc)^2 + (mc^2)^2} - mc^2$$

Substituting equation (50) in the above equation provides:-

$$K = mc^2\left[\sqrt{1 + \frac{n^2h^2}{4m^2L^2c^2}} - 1\right] \tag{58}$$

**Equation (58) is the energy of a particle moving with relativistic velocity, trapped in a box.** If the quantity $n^2\ h^2/4m^2L^2c^2$ is very less than unity, then equation (58) transforms to the non-relativistic energy expression, by the binomial expansion, given as [64-73]:-

$$K_n = \frac{n^2h^2}{8mL^2}$$

The satisfactory conversion of the relativistic energy Eigen value expression into the non relativistic energy expression (58), demonstrates its correctness.

### 22.2 The general case

From the general kinetic energy-momentum equation (42):-

$$K = P\upsilon_p$$

Substituting equation (57) in above equation offers:-

$$K = \frac{nh}{2L}(\upsilon_p) \tag{59}$$

Equation (59) presents the general energy expression for the particle trapped in a box. **The correctness of equation (59) can be verified by the following two ways:-**

**First**, convert eq. (59) into eq. (58), for the relativistic cases

To completely specify equation (59), substitution of the phase velocity term in it, is necessary. For that, the velocity of particle should be specified. Equation (57) offers the momentum of the particle; hence velocity is determinable from it, i.e. eq. (57) can be written as:-

$$\frac{m\upsilon}{\sqrt{1-\upsilon^2/c^2}} = \frac{nh}{2L}$$

So, the velocity of body becomes:-

$$\upsilon = \left[\frac{nh}{2mL}\right] \times \frac{1}{\sqrt{1 + \left(\frac{nh}{2mL}\right)^2 \frac{1}{c^2}}} \tag{60}$$

As from eq. (20):-

$$\upsilon_p = \frac{c^2}{\upsilon}\left[1 - \sqrt{1 - \left(\frac{\upsilon}{c}\right)^2}\right]$$

Substituting equation (60) in above equation provides:-

$$\upsilon_p = \frac{2mLc^2}{nh}\left[\sqrt{1 + \frac{n^2h^2}{4m^2L^2c^2}} - 1\right] \tag{61}$$

Substituting equation (61) in equation (59) gives:-

$$K = mc^2\left[\sqrt{1 + \frac{n^2h^2}{4m^2L^2c^2}} - 1\right] \tag{62}$$

Energy expression for the particle thus obtained is identically same as that obtained in equation (58). Therefore, the expression $K = mc^2\left[\sqrt{1 + \frac{n^2h^2}{4m^2L^2c^2}} - 1\right]$, for relativistic cases, is merely an expanded form of the general energy expression:

$$K = \frac{nh\upsilon_P}{2L}$$

**Second** technique to verify the general energy expression [eq.(59)] is its reduction into the non-relativistic energy expression, $n^2h^2/8mL^2$ . As, for non-relativistic cases, phase velocity becomes *υ/2*, [eq. (22)]. Therefore, equation (59) for non-relativistic cases becomes:-

$$K' = \frac{nh\upsilon}{4L}$$

$$\Rightarrow K' = \frac{nh(m\upsilon)}{4L(m)} = \frac{nhP}{4mL} \qquad (63)$$

Substituting the momentum of the particle [eq. (57)] in equation (63) provides:-

$$K' = \frac{n^2h^2}{8mL^2}$$

The above expression presents identically the same energy expression for the particle trapped in a box, moving at non-relativistic velocity [64-73]. The conversion of relativistic energy equation, $nh\upsilon_p/2L$ to non-relativistic energy equation $n^2h^2/8mL^2$ shows the appropriateness of general energy relation, $K = \frac{nh\upsilon_p}{2L}$.

## 23. INFINITE POTENTIAL WELL SITUATION BY APPLICATION OF THE PROPOSED WAVE EQUATIONS – GENERAL AND RELATIVISTIC

The theoretical situation of a particle trapped in an infinite potential well is now tackled by the proposed wave equations, to determine the energy Eigen values (for general as well as relativistic cases) and the wave function of the particle trapped in it. Consider a particle of inertial mass *m*, trapped in one dimensional potential specified as:-

$$V(x)\begin{cases}\infty;(x<0,x>L)\\0;(0<x<L)\end{cases}$$

As the energy of particle can never be infinite, thus it does not exist outside the well, as represented in figure 6. Therefore its wave function is zero, for $x \le 0$ and $x \ge L$.

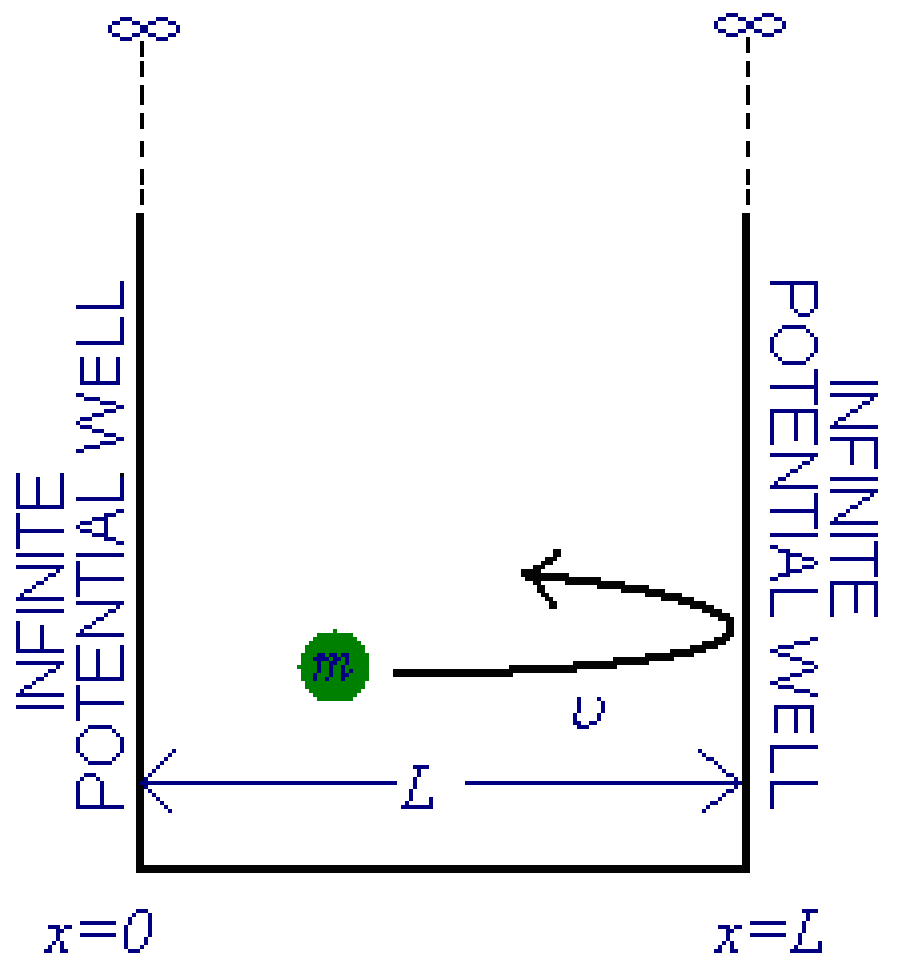

**Figure 6. Particle Trapped In an Infinite Potential Well**

The task is to obtain the wave function $\psi$ and the energy Eigen values of the particle within the boundaries of the well.

### 23.1 The general case

The application of the general wave equation (48) in the situation determines the particle's general behavior, applicable for both relativistic and non-relativistic cases. Since the potential within the box is *0*, the general quantum mechanical wave equation (48), for $0 \le x \le L$, can be written as:-

$$\frac{d^2\psi}{dx^2} + \frac{E^2}{\hbar^2\upsilon_p^2}\psi = 0 \qquad (64)$$

Solution of the above differential equations is of the form:-

$$\psi = A\sin\left((\frac{E}{\hbar\upsilon_p})x\right) + B\cos\left((\frac{E}{\hbar\upsilon_p})x\right)$$

where, *A* and *B* are constants.

***Applying Boundary Conditions***

1) At $x=0; \psi=0$.

Cosine term is not zero at $x=0$, hence $B$=0.

$$\Rightarrow \psi = [A\sin(\frac{Ex}{\hbar\upsilon_p})] \tag{65}$$

2) At $x=L$, $\psi=0$

$$\psi\big|_{x=L} = \left[ A\sin(\frac{EL}{\hbar\upsilon_p}) \right]$$

$$\Rightarrow 0 = \left[ A\sin(\frac{EL}{\hbar\upsilon_p}) \right]$$

$$\Rightarrow \frac{EL}{\hbar\upsilon_p} = n\pi \tag{66}$$

$$E_n = \frac{nh\upsilon_p}{2L} \tag{67}$$

The above expression (67), for the energy of a particle is identical to the one obtained in the previous section, i.e. energy of the trapped particle is an integral multiple of $h\upsilon_p/2L$.

Moreover, from equation (66):-

$$\frac{EL}{\hbar\upsilon_p} = n\pi \tag{68}$$

Since $U=0$, within the box, the total energy ($E=K+U$) of the particle is the kinetic energy, i.e. within the box we have:-

$$E=K$$

From equation (42):-

$$K = P\upsilon_P$$

Therefore, within the box,

$$E = P\upsilon_p \tag{69}$$

Comparing, equation (68) and (69) we get:-

$$P = \frac{nh}{2L}$$

Substituting $P = \frac{h}{\lambda}$ in above equation gives:-

$$L = \frac{n\lambda}{2}$$

The matter waves of the particle forms a standing wave pattern between the walls of the box, irrespective of its velocity.

Substituting eq. (68) in eq. (65), wave function $\psi$ becomes:-

$$\psi_n = \left[ A\sin(\frac{n\pi x}{L}) \right]$$

***Normalization of $\psi$***

The integral of $|\psi_n|^2$ inside the box, should be unity.

So, the constant A becomes equal to $\sqrt{2/L}$.

The wave function becomes:-

$$\psi_n = \left[ \sqrt{\frac{2}{L}} \sin\left(\frac{n\pi x}{L}\right) \right] \tag{70}$$

The obtained wave function is precisely the same wave function of the particle, obtained by employing Schrödinger's equation [28,66,69]. This shows the accuracy of the proposed general quantum mechanical wave equation.

### 23.2 The relativistic case

The relativistic wave equation (51) for the situation ($0<x<L$) can be written as:-

$$\frac{d^2\psi}{dx^2} + \alpha^2\psi = 0 \tag{71}$$

where, $\alpha = \sqrt{\frac{2mE}{\hbar^2} + \frac{E^2}{(\hbar c)^2}}$ .Solving equation (71) and applying the boundary conditions, provides the energy Eigen values and wave-function of the particle to be:-

$$E_n = mc^2\left(\sqrt{1 + \frac{n^2h^2}{4m^2c^2L^2}} - 1\right)$$

$$\psi_n = \left[\sqrt{\frac{2}{L}}\sin\left(\frac{n\pi x}{L}\right)\right]$$

The energy Eigen value and the wave function are identical to that obtained in equation (62) and (70). Therefore, the results obtained are well justified. From this section it is concluded that the general and the relativistic energy expression of a particle trapped in an infinite potential well, is:-

$$E_n = \frac{nh\upsilon_p}{2L} \text{ and } E_n = mc^2\left(\sqrt{1+\frac{n^2h^2}{4m^2c^2L^2}}-1\right)$$

## 24. BARRIER PENETRATION-RELATIVISTIC

The situation of a relativistic particle having energy '*E*' encountering a potential barrier of height '*U*', where *E*<<*U*, is discussed in this section. The situation is illustrated in figure 7.

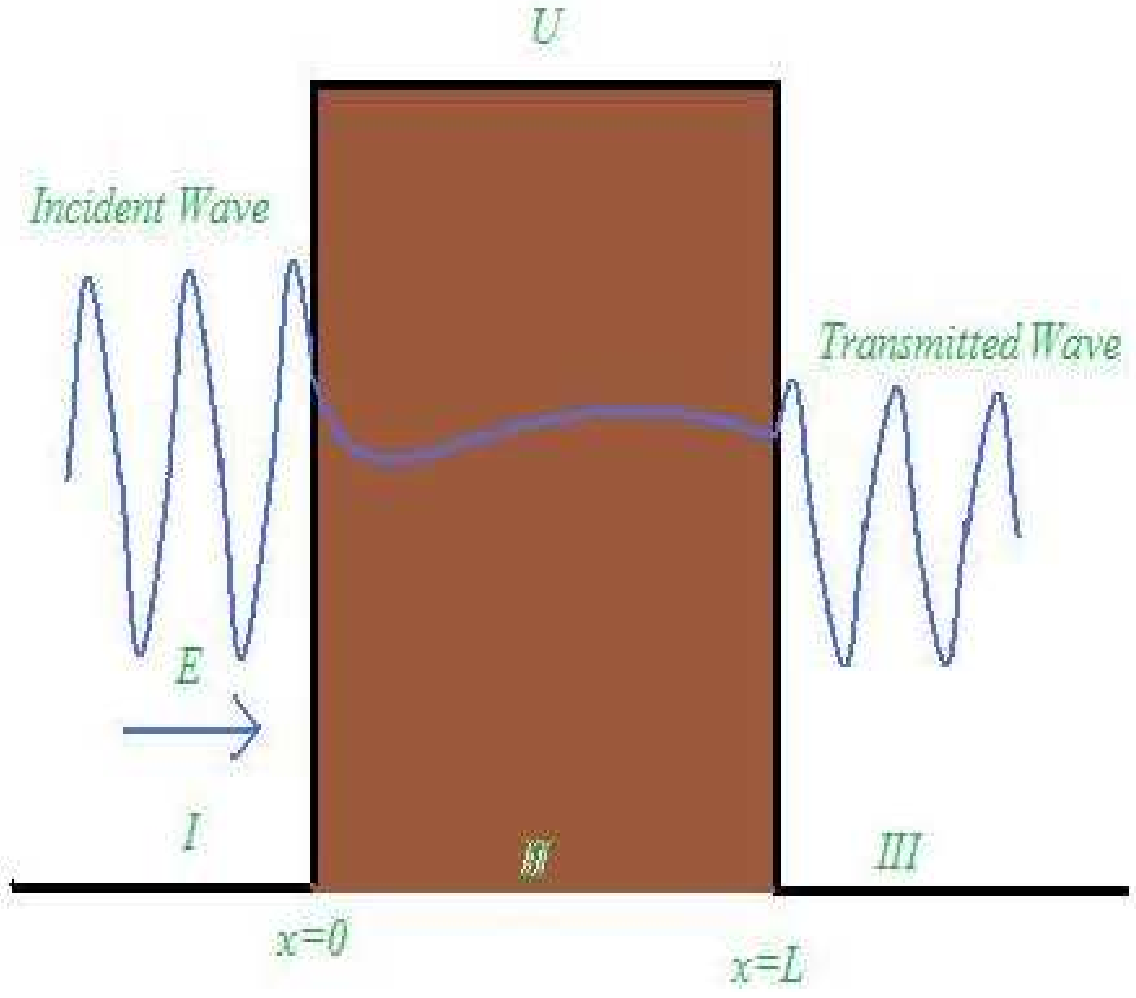

**Figure 7 Particle of Energy *E*<<*U*, incident on a Potential Barrier of height *A*nd *L* wide.**

For region I, potential acting on the particle is zero. Therefore, the relativistic wave equation (51) for the region I take the form:-

$$\frac{d^2\psi_1}{dx^2} + \rho^2\psi_1 = 0 \qquad (72)$$

$$\rho = \sqrt{\frac{2mE}{\hbar^2} + \frac{E^2}{\hbar^2c^2}}$$

When the beam of particles encounters the barrier at *x=0*, some of them gets reflected back and the other gets transmitted to region II. On entering the region II, particle's velocity decreases severely because the potential acted on particle is much greater than its incident Kinetic energy *E*. As a result in region II, the velocity becomes non-relativistic and **thus only Schrodinger's equation is applicable for this region**:-

$$\frac{d^2\psi_2}{dx^2} - \eta^2\psi_2 = 0 \quad ;\eta^2 = \frac{2m(U-E)}{\hbar^2} \qquad (73)$$

Similarly, for region III relativistic wave equation becomes:-

$$\frac{d^2\psi_3}{dx^2} + \rho^2\psi_3 = 0 \qquad (74)$$

The solution of the wave equations (72), (73) and (74) are of the form:-

$$\psi_1 = A\exp(i\rho x) + B\exp(-i\rho x)$$
$$\psi_2 = C\exp(\eta x) + D\exp(-\eta x)$$
$$\psi_3 = F\exp(i\rho x) + G\exp(-i\rho x)$$

The transmission probability *T* for a particle to pass through the barrier is to be calculated. Transmission probability is defined as the ratio of the flux of particle emerging from the barrier and the flux that arriving it [28,66]. Application of boundary conditions at *x=0* and *x=L*, similar to that followed for non-relativistic cases [28,66,58,68], provides us the transmission probability to be:-

$$T = \left[16\frac{E}{U}\left(1-\frac{E}{U}\right)\exp(-2\eta L)\right]\frac{\left[1+\frac{E}{2mc^2}\right]}{\left[1+\frac{E}{2mc^2}\times\frac{E}{U}\right]^2}$$

Since *U*>>*E*, therefore, an approximate transmission probability for a particle encountering the rectangular barrier of height *U* and length *L*, at a relativistic energy *E* is:-

$$T \cong \frac{\left[1+\frac{E}{2mc^2}\right]}{\left[1+\frac{E}{2mc^2}\times\frac{E}{U}\right]^2}\times\exp(-2\eta L) \quad (75)$$

The above transmission probability expression for the non-relativistic cases ($E<<2mc^2$) becomes:-

$$T = \exp(-2\eta L)$$

which is the transmission probability for non-relativistic cases [28,66]. Therefore, **the relativistic transmission probability expression (75) reduces to the non-relativistic expression, exactly.**

## 25. GENERAL QUANTUM HARMONIC OSCILLATOR: USING UNCERTAINITY PRINCIPLE

In this section, the general case of a particle oscillating simple harmonically is dealt, applicable for both non-relativistic and relativistic velocity of body. The potential acting on particle is:-

$$U = \frac{m\omega^2 x^2}{2}$$

The application of uncertainty principle in non-relativistic situation provides us minimum energy, which a particle possesses even in its lowest possible state. This minimum energy is given by [63-73,85-87]:

$$E_{\min} = \frac{\hbar\omega}{2}$$

This section deals with the same situation, under the assumption that the velocity of the particle may be relativistic or non-relativistic, in its lowest energy state. Therefore, a general minimum energy expression for such situations obtained, reducible to the non-relativistic minimum energy expression as well. The non-zero energy phenomenon in quantum harmonic oscillator is a consequence of the uncertainty principle [66]. Therefore, the first approach towards this problem is to use the uncertainty principle. The total energy of a particle is the sum of the kinetic energy ($K$) and the potential energy ($U$) acting on it. i.e.:-

$$E = K + U$$

The use of the relativistic kinetic energy expression $K = \sqrt{(pc)^2 + (mc^2)^2} - mc^2$, in above equation leads mathematics on an extremely tedious way. Therefore, use of the kinetic energy-momentum relation [eq. (42)] is preferred. So that the total energy of the particle becomes:-

$$\langle E\rangle = \langle P\rangle \upsilon_p + \frac{m\omega^2\langle x\rangle^2}{2} \quad (76)$$

Uncertainty principle, under extreme accuracy condition offers [88,89]:-

$$\Rightarrow \Delta p = \frac{\hbar}{2\Delta x}$$

Substituting above relation in equation (76):-

$$E = \frac{\hbar\upsilon_p}{2\Delta x} + \frac{m\omega^2(\Delta x)^2}{2} \quad (77)$$

To get minimum value of energy the derivative of $E$lis substituted equal to zero, i.e.:-

$$\frac{dE}{d\Delta x} = 0$$

$$\Rightarrow -\frac{\hbar\upsilon_p}{2(\Delta x)^2} + \frac{m\omega^2\Delta x}{1} = 0$$

$$\Rightarrow \Delta x = [\frac{\hbar\upsilon_p}{2m\omega^2}]^{1/3} \quad (78)$$

Equation (78) provides the minimum uncertainty in the position of the particle. The corresponding uncertainty in momentum is:-

$$\Delta P = \left[\frac{\hbar^2 m\omega^2}{4\upsilon_p}\right]^{1/3} \quad (79)$$

Substituting eq. (78) in eq. (77), the general minimum energy expression is found to be:-

$$E_{\min} = 0.94(m\omega^2\hbar^2\upsilon_p^2)^{1/3} \approx \left(m\omega^2\hbar^2\upsilon_p^2\right)^{1/3} \quad (80)$$

Above expression is the proposed general minimum energy expression, for a quantum harmonic oscillator, valid for both relativistic and non-relativistic cases. To completely specify the minimum energy expression substitution of the phase velocity ($\upsilon_p$), in equation (80) is necessary. Equation (79) can be used to determine the velocity of body and thereafter the phase velocity. First verification of the minimum energy expression {eq. (80)} is to ensure its consistency with the non-relativistic minimum energy expression $E_{\min} = \frac{\hbar\omega}{2}$. As for non-relativistic cases [from eq. (22)]:-

$$\Delta P = m\Delta\upsilon \text{ and } \upsilon_p = \frac{\Delta\upsilon}{2}$$

Thus eq. (79) and (80), for non-relativistic cases becomes:-

$$m\Delta\upsilon = \left[\frac{m\omega^2\hbar^2}{2\Delta\upsilon}\right]^{1/3} \quad (81)$$

$$E_{\min} = \left(m\omega^2\hbar^2\frac{(\Delta\upsilon)^2}{4}\right)^{1/3} \quad (82)$$

Solving equation (81) gives:

$$(\Delta\upsilon)^2 = \frac{\hbar\omega}{\sqrt{2}m} \quad (83)$$

Substitution of eq. (83) in (82) gives:-

$$E_{\min} \approx \frac{\hbar\omega}{2}$$

which is the minimum energy of a Quantum Harmonic Oscillator, for non-relativistic cases [63-73,85-87].Therefore, from this section it is concluded that the general minimum energy expression for the quantum harmonic oscillator is:-

$$E_{\min} = \left(m\omega^2\hbar^2\upsilon_p^2\right)^{1/3}$$

## 26. QUANTUM HARMONIC OSCILLATOR BY THE APPLICATION OF THE PROPOSED GENERAL WAVE EQUATION

The general situation of quantum harmonic oscillator is now dealt with the proposed general wave equation (48). The general Wave Equation is:-

$$\frac{d^2\psi}{dx^2} + [\frac{E-U}{\hbar\upsilon_P}]^2\psi = 0 \quad (84)$$

As the potential acting on the particle performing simple harmonic oscillation is $U = \frac{m\omega^2x^2}{2}$. Therefore, equation (84) becomes:-

$$\frac{d^2\psi}{dx^2} + [\frac{E}{\hbar\upsilon_P} - \frac{m\omega^2x^2}{2\hbar\upsilon_p}]^2\psi = 0$$

$$\frac{d^2\psi}{dx^2} + \left(\frac{E^2}{\hbar^2\upsilon_p^2}\right)\psi + \left(\frac{m^2\omega^4x^4}{4\hbar^2\upsilon_p^2}\right)\psi - \left(\frac{Em\omega^2x^2}{\hbar^2\upsilon_p^2}\right)\psi = 0$$

(85)

Assume $\psi = C\exp(-\frac{\alpha x^2}{2})$, $\alpha$ is constant, because for $x \to \pm\infty, \psi \to 0$. Hence eq. (85) becomes:-

$$\left(\alpha^2 - \frac{Em\omega^2}{\hbar^2\upsilon_p^2}\right)x^2 + \left(\alpha - \frac{E^2}{\hbar^2\upsilon_p^2}\right) + \left(\frac{m^2\omega^2}{4\hbar^2\upsilon_p^2}\right)x^4 = 0$$

Comparing both sides of above equation provides:-

$$\alpha^2 - \frac{Em\omega^2}{\hbar^2\upsilon_p^2} = 0; \alpha - \frac{E^2}{\hbar^2\upsilon_P^2} = 0; \left(\frac{m\omega}{2\hbar\upsilon_P}\right)^2 = 0$$

The last equation is neglected because it suggests phase velocity to be infinite; which is not possible [eq. (20)]. Therefore, solving other two equations provides:-

$$E_{\min} = \left(m\omega^2\hbar^2\upsilon_p^2\right)^{1/3}$$

This is the general minimum energy expression of a simple harmonic oscillator. The expression thus obtained by solving the proposed general quantum mechanical wave equation is identical to that obtained in previous section, using the uncertainty principle. By using the energy expression ($E$) the value of constant α becomes:-

$$\alpha = \frac{E^2}{\hbar^2 \upsilon_P^2} = \frac{(m\omega^2 \hbar^2 \upsilon_P^2)^{2/3}}{\hbar^2 \upsilon_P^2} = \left(\frac{m\omega^2}{\hbar \upsilon_P}\right)^{\frac{2}{3}}$$

Therefore, the general wave function of the particle in its lowest energy state becomes:-

$$\psi = C \exp\left[-\left(\frac{m\omega^2}{\hbar \upsilon_P}\right)^{\frac{2}{3}} \frac{x^2}{2}\right]$$

## 27. PARTICLE MOVING IN A SPHERICALLY SYMMETRIC POTENTIAL- GENERAL AND RELATIVISTIC

In this section, the situation consisting of a particle acted by a spherically symmetric potential is dealt. The general and relativistic wave equations are obtained.

### 27.1 The general case

The general wave equation (48) is utilized in this situation. In spherical polar coordinates the equation becomes:-

$$\frac{1}{r^2}\frac{\partial}{\partial r}\left(r^2 \frac{\partial \psi}{\partial r}\right) + \frac{1}{r^2 \sin\theta}\frac{\partial}{\partial \theta}\left(\sin\theta \frac{\partial \psi}{\partial \theta}\right) + \frac{1}{r^2 \sin^2\theta}\frac{\partial^2 \psi}{\partial \phi^2} + \left(\frac{E-U}{\hbar \upsilon_P}\right)^2 \psi = 0 \qquad (86)$$

Equation (86) is then separable into three individual equations by writing [28,66,69,73]:-

$$\psi(r,\theta,\phi) = R(r)\Theta(\theta)\Phi(\phi)$$

and multiplying the whole equation by $\frac{r^2 \sin^2\theta}{R\Theta\Phi}$. So that equations (86) separates into following equations:-

**Equation for Φ**: $\frac{d^2\Phi}{d\phi^2} + m_l^2 \Phi = 0$

*The Φ equation obtained from the proposed general wave equations is identical to the non-relativistic equation* [66,69,73] and is independent of phase velocity term. Thus a particle, even if moving at relativistic velocity, follows the same Φ equation, as that followed by non-relativistic particle. The term $m_l$ is a constant, already called in literature as *magnetic quantum number*. The solution of this equation is already known in literature [28,66,73]. The permissible values of $m_l$ are:-

$$m_l = 0, \pm 1, \pm 2, .....$$

**Equation for Θ:**

$$\frac{1}{\sin\theta}\frac{d}{d\theta}\left(\sin\theta \frac{d\Theta}{d\theta}\right) + \left[l(l+1) - \frac{m_l^2}{\sin^2\theta}\right]\Theta = 0 \qquad (87)$$

*The Θ equation is also the same as that for the non-relativistic cases* [29,66,73]. The term $l$ is defined as orbital quantum number and its permissible values of $l$ are:-

$$l = 0,1,2,3.....$$

**Equation for R:**

$$\frac{1}{r^2}\frac{d}{dr}\left(r^2 \frac{dR}{dr}\right) + \left[\left(\frac{E-U}{\hbar \upsilon_P}\right)^2 - \frac{l(l+1)}{r^2}\right]R = 0 \qquad (88)$$

Equation (88) is the general radial wave equation. This equation can describe the behavior of body moving in a spherically symmetric potential, both for the relativistic and non-relativistic cases.

The task of separating the wave equation (86) is accomplished and we obtain three individual wave equations. Out of these equations, the Φ and Θ equations are independent of the velocity term and identically same to the non-relativistic equations. However, the radial wave equation (88) is dependent on the velocity term. Therefore, **the particle's behavior, at relativistic velocity, differs from the non-relativistic behavior only in radial aspect**. The radial wave equation can be converted into the non-relativistic radial equation as follows:-

Since, $(E-U)=K=P\upsilon_P$ from equation (22). Therefore, equation becomes:-

$$\frac{1}{r^2}\frac{d}{dr}\left(r^2\frac{dR}{dr}\right)+\left[\frac{P^2}{\hbar^2}-\frac{l(l+1)}{r^2}\right]R=0$$

For non-relativistic case, $P^2=2mK=2m(E-U)$:-

$$\Rightarrow\frac{1}{r^2}\frac{d}{dr}\left(r^2\frac{dR}{dr}\right)+\left[\frac{2m(E-U)}{\hbar^2}-\frac{l(l+1)}{r^2}\right]R=0$$

The radial wave equation thus obtained is identical to the non-relativistic radial wave equation [28,66,85]. Similarly, the radial wave equation for the relativistic situations can also be obtained from the general radial wave equation (88). In the subsequent section, it is derived separately by utilizing the relativistic wave equation (51).

### 27.2 The relativistic case

The proposed relativistic wave equation (51) for the spherically symmetric potential segregates into three individual wave equations; out of which the Φ and Θ equation are the same as proposed previously. However, **the radial wave equation, for the relativistic cases** becomes:-

$$\frac{1}{r^2}\frac{d}{dr}\left(r^2\frac{dR}{dr}\right)+\left[\left(\frac{2m(E-U)}{\hbar^2}+\frac{(E-U)^2}{\hbar^2c^2}\right)-\frac{l(l+1)}{r^2}\right]R=0 \quad (89)$$

For the solution of the radial equation, the explicit form of the potential energy function $U$ is an essential requirement. In the upcoming section, the situation of Rigid Rotor is dealt with the proposed general and relativistic set of wave equations.

## 28. RIGID ROTOR-GENERAL AND RELATIVISTIC

A rigid rotor consists of two masses $m_1$and $m_2$ separated by a fixed distance $r$, so that the potential energy acting on the system is zero [66,69,85]. The rotation of the system is considered only about the axis passing through the centre of mass and perpendicular to the plane containing the two masses. Due to the fixed separation distance ($r$) between two masses, the wave function depends only on the angles $\theta$ and $\phi$. The general and relativistic behavior of the system is investigated in this section.

### 28.1 The general case

The proposed general wave equation (48), in spherical coordinates (equation 86), for the rigid rotor, becomes:-

$$\frac{1}{r^2\sin\theta}\frac{\partial}{\partial\theta}\left(\sin\theta\frac{\partial\psi}{\partial\theta}\right)+\frac{1}{r^2\sin^2\theta}\frac{\partial^2\psi}{\partial\phi^2}+\frac{E^2}{(\hbar\upsilon_P)^2}\psi=0$$

$$\Rightarrow\frac{1}{\sin\theta}\frac{\partial}{\partial\theta}\left(\sin\theta\frac{\partial\psi}{\partial\theta}\right)+\frac{1}{\sin^2\theta}\frac{\partial^2\psi}{\partial\phi^2}+\frac{E^2r^2}{(\hbar\upsilon_P)^2}\psi=0$$

Since the system is independent of the radial aspect (due to the fixed separation distance), the initial term of equation (86) is absent. Writing $\psi(\theta,\phi)=\Theta(\theta)\Phi(\phi)$ produces following two equations:-

$$\frac{d^2\Phi}{d\phi^2}+m_l^2\Phi=0 \quad (90)$$

$$\frac{1}{\sin\theta}\frac{d}{d\theta}\left(\sin\theta\frac{d\Theta}{d\theta}\right)+\left[\frac{E^2r^2}{\hbar^2\upsilon_P^2}-\frac{m_l^2}{\sin^2\theta}\right]\Theta=0 \quad (91)$$

The solution of equation (90) is already given in section 17.1. However, the solution of equation (91) involves the introduction of a new variable *z=cos(θ)*, so that equation becomes a Legendre equation [66]. Therefore, we obtain the condition:-

$$\frac{(Er)^2}{(\hbar\upsilon_P)^2}=l(l+1)$$

$$\Rightarrow E\times r=\sqrt{[l(l+1)]}\hbar\upsilon_P \quad (92)$$

As, the potential acting on the particle is zero, $E$ (sum of Kinetic and potential energy) becomes equal to the Kinetic energy. Therefore, equation (92) becomes:-

$$K\times r=\sqrt{l(l+1)}\hbar\upsilon_P$$

Substituting $K = Pv_P$ from equation (42), in above equation provides:-

$$rP = \sqrt{l(l+1)}\hbar \qquad (93)$$

Angular momentum ($L$) of a rotating body is defined as:-

$$L = \vec{r} \times \vec{P} = rP \sin\chi$$

where, $\chi$ is the angle between the radius and momentum vector. As the rotor rotates about the fixed axis passing through the centre of mass perpendicular to the plane containing the two masses, $\chi$ becomes equal to $90^0$. Therefore, the quantity $rP$ in equation (93) represents the angular momentum of the rigid rotor:-

$$\Rightarrow L = \sqrt{l(l+1)}\hbar \qquad (94)$$

As $L$ is independent of the velocity of body, **the angular momentum of a rigid rotor follows only this unique quantization rules, irrespective of its velocity.** Equation (94) is valid for both non-relativistic and relativistic cases. From the angular momentum expression (94), the energy of the rigid rotor is determinable. As, for the non-relativistic cases, the kinetic energy of the rotor is related to the angular momentum by the relation:-

$$K = \frac{L^2}{2I} \qquad (95)$$

where, $I$ is the moment of inertia, defined as:-

$$I = \mu r^2 \; ; \mu = \frac{m_1 m_2}{m_1 + m_2}$$

The quantity $\mu$ is defined as the reduced mass of the system [66]. Therefore, the energy of rotor for non-relativistic cases, by the substitution of equation (94) in expression (95) provides:-

$$K = l(l+1)\frac{\hbar^2}{2I}$$

The obtained energy expression for non-relativistic cases is identical to that obtained by the application of the Schrödinger's equation [66,85]. Therefore, the angular momentum expression (94) is justified.

Similarly, to obtain the energy of rigid rotor for relativistic cases, a relativistic relation between Kinetic energy and momentum is required. From relativity [28]:-

$$K = \sqrt{\mu^2 c^4 + (p^2 c^2)} - \mu c^2 = \mu c^2 \left[ \sqrt{1 + \frac{P^2}{\mu^2 c^2}} - 1 \right]$$

$$\Rightarrow K = \mu c^2 \left[ \sqrt{1 + \frac{r^2 P^2}{(\mu r^2)\mu c^2}} - 1 \right]$$

$$\Rightarrow K = \mu c^2 \left[ \sqrt{1 + \frac{L^2}{(I)\mu c^2}} - 1 \right]$$

Substituting the equation (94) in above equation gives:-

$$K = \mu c^2 \left[ \sqrt{1 + \frac{l(l+1)\hbar^2}{(I)\mu c^2}} - 1 \right] \qquad (96)$$

**Above expression provides the energy of a rigid rotor rotating at relativistic velocity.**

If the quantity $\frac{l(l+1)\hbar^2}{I\mu c^2} << 1$, then the energy expression (96) becomes:-

$$K = l(l+1)\frac{\hbar^2}{2I}$$

which is the energy of the rotor for non-relativistic cases. Thus the relativistic energy expression (96) is well justified.

### 28.2 The relativistic case

The situation of relativistic rigid rotor is now dealt with the help of relativistic wave equation (89). The equation (89) for rigid rotor becomes:-

$$\frac{1}{\sin\theta}\frac{\partial}{\partial\theta}\left(\sin\theta\frac{\partial\psi}{\partial\theta}\right) + \frac{1}{\sin^2\theta}\frac{\partial^2\psi}{\partial\phi^2} + r^2\left[\frac{2\mu E}{\hbar^2} + \left(\frac{E}{\hbar c}\right)^2\right]\psi = 0$$

Separating the above equation and solving the $\Theta$ equation provides:-

$$E = \pm \mu c^2 \left[ \sqrt{1 + \frac{l(l+1)\hbar^2}{(I)\mu c^2}} - 1 \right]$$

As, $E$ is equal to kinetic energy (due to the absence of the potential energy function $U$), negative energy solution is neglected. Therefore:-

$$E = \mu c^2 \left[ \sqrt{1 + \frac{l(l+1)\hbar^2}{(I)\mu c^2}} - 1 \right]$$

The energy expression thus obtained is identical to equation (96). Therefore, the obtained result is justified. In the present paper the situation of rigid rotor is the last application of the proposed quantum mechanical equations. The equations are further applicable to other bound state situations also.

## 29. CONCLUSION

The aim of the paper is not to reject or cancel out the theory of de Broglie, which has presented its significance in the birth of quantum physics, rather to propose more general and completely correct formulations. The modified frequency and phase velocity expression for matter waves are free from all the inadequacies confronted by de Broglie's proposed expressions. The proofs, presented in the paper, regarding the proposed modified expressions for matter waves shows the correctness of the work, beyond any doubt. Therefore, the widely accepted relation $h\nu=\gamma mc^2$ is true for photons, not for matter waves. On the other hand the frequency expression $h\nu=(\gamma-1)mc^2$ proposed in the paper is applicable for both photon and matter. The dependency of particle type characteristic on the wave characteristic suggests that every wave should possess momentum and kinetic energy. The concepts proposed in the paper are applied to introduce general and relativistic Quantum Mechanical Wave Equations and their applications in a variety of conditions confirmed their correctness. The general wave equations determine the general behavior of particle, applicable for both relativistic and non-relativistic cases. At the same time, the relativistic wave equation determines the behavior of particle at relativistic velocity. The energy Eigen values obtained for the general and the relativistic cases can be well transformed into the non-relativistic expression. Therefore, satisfactory results are obtained by their applications. The usefulness of the proposed concepts is associated with the relativistic aspects of quantum physics. Consequently, introduction of these concepts within the domain of Quantum Physics is essential.

## ACKNOWLEDMENTS

Authors gratefully acknowledges the initiatives and support from TIFAC-Center of Relevance and Excellence in Fiber Optics and Optical Communication at Delhi College of Engineering, now DTU, Delhi under Mission REACH program of Technology Vision-2020, Government of India. Authors are also thankful to Mr. Saumil Joshi, Research Student, University of Colorado at Boulder, for providing the soft copies of historical research papers in this area. One of the authors (RKS) acknowledges (*i*) His teacher Prof. Ajoy K. Ghatak (IIT Delhi) for creating interest in teaching of Basic Quantum Mechanics (*ii*) all his past and present undergraduate students for lively discussion related to this topic.